\documentclass[12pt]{article}
\usepackage{epsfig, amsmath, float, graphicx, color}
\newcommand{\ltsim}{ \mathop{}_{\textstyle \sim}^{\textstyle <} }
\newcommand{\lsim}{ \mathop{}_{\textstyle \sim}^{\textstyle <} }

\newcommand{\gtsim}{ \mathop{}_{\textstyle \sim}^{\textstyle >} }
\newcommand{\gsim}{ \mathop{}_{\textstyle \sim}^{\textstyle >} }

\def\SGone{S\,G\,\,1} \def\SGtwo{S\,G\,\,2}
\def\GMone{GM\,1}     \def\GMtwo{GM\,2}

\def\Im{\mathop{\rm Im}\nolimits}

\def\alps{\alpha_s}
\def\ddelg{\Del\ov{\delta}_G^{}}
\def\Del{\Delta}
       
\def\dgfa{\Del g_\alpha^{f}}
\def\dgzbarsq{\Del \gzbarsq}
\def\dmw{\Del m_W^{}}

\def\ds{\Del S}    \def\dt{\Del T}    \def\du{\Del U}    
\def\dsz{\Del S_Z} \def\dtz{\Del T_Z} 
\def\eff{{\rm eff}}

\def\gzbarsq{\bar{g}_Z^2}
\def\hph{\hphantom{-}} \def\hpz{\hphantom{0}} \def\hpzz{\hphantom{00}}
\def\mh{m_{H_{\rm SM}}}

\def\mt{m_t^{}}
\def\mw{m_W}

\def\mz{m_Z}
\def\mzsq{m_Z^2}
\def\ov{\overline}
\def\sbarsq{\bar{s}^2} \def\dsbarsq{\Del \sbarsq}
\def\simgt{\,{\rlap{\lower 3.5pt\hbox{$\mathchar\sim$}}\raise 1pt\hbox{$>$}}\,}
\def\simlt{\,{\rlap{\lower 3.5pt\hbox{$\mathchar\sim$}}\raise 1pt\hbox{$<$}}\,}
\def\smr{{\rm SM}}
\def\sbottom{\wt{b}}

\def\stop{\wt{t}}

\def\sz{S_Z}

\def\wt{\widetilde}
\def\xa{x_\alpha^{}}
\def\xh{x_h}   %\def\xh{x_h^{}}
\def\xs{x_s}   %\def\xs{x_s^{}}
\def\xt{x_t}   %\def\xt{x_t^{}}

\newcommand{\bea}{\begin{eqnarray}}
\newcommand{\eea}{\end{eqnarray}}
\newcommand{\bsub}{\begin{subequations}}
\newcommand{\esub}{\end{subequations} \noindent}

\makeatletter
\newtoks\@stequation

\def\subequations{\refstepcounter{equation}%
  \edef\@savedequation{\the\c@equation}%
  \@stequation=\expandafter{\theequation}%   %only want \theequation
  \edef\@savedtheequation{\the\@stequation}% %expanded once
  \edef\oldtheequation{\theequation}%
  \setcounter{equation}{0}%
  \def\theequation{\oldtheequation\alph{equation}}}

\def\endsubequations{%
  \ifnum\c@equation < 2 \@warning{Only \the\c@equation\space subequation
    used in equation \@savedequation}\fi
  \setcounter{equation}{\@savedequation}%
  \@stequation=\expandafter{\@savedtheequation}%
  \edef\theequation{\the\@stequation}%
  \global\@ignoretrue}

\def\eqnarray{\stepcounter{equation}\let\@currentlabel\theequation
\global\@eqnswtrue\m@th
\global\@eqcnt\z@\tabskip\@centering\let\\\@eqncr
$$\halign to\displaywidth\bgroup\@eqnsel\hskip\@centering
     $\displaystyle\tabskip\z@{##}$&\global\@eqcnt\@ne
      \hfil$\;{##}\;$\hfil
     &\global\@eqcnt\tw@ $\displaystyle\tabskip\z@{##}$\hfil
   \tabskip\@centering&\llap{##}\tabskip\z@\cr}

\makeatother

\begin{document}

\titlepage

\begin{flushright}
KEK-TH-1236\\
OCHA-PP-298
\end{flushright}

\vspace*{1.5cm}

\begin{center}

{\Large \bf The MSSM confronts the precision electroweak
            data and the muon $g-2$}

\vspace*{1cm} {\bf Gi-Chol Cho}$^a$, {\bf Kaoru Hagiwara}$^{b,c}$, 
{\bf Yu Matsumoto}$^{a,b}$\\ and {\bf Daisuke Nomura}$^d$ \\

\vspace*{0.5cm}

$^a$ {\em Department of Physics, Ochanomizu University, 
          Tokyo 112-8610, Japan}\\
$^b$ {\em KEK Theory Center, Tsukuba 305-0801, Japan}\\
$^c$ {\em Sokendai,          Tsukuba 305-0801, Japan}\\
$^d$ {\em Department of Physics, Tohoku University,
           Sendai 980-8578, Japan}
\end{center}

\vspace*{0.5cm}

\begin{abstract}
\noindent
We update the electroweak study of the predictions of the Minimal
Supersymmetric Standard Model (MSSM) including the
recent results on the muon anomalous magnetic moment, the
weak boson masses, and the final precision data on the $Z$
boson parameters from LEP and SLC.  
We find that the region of the parameter space where the slepton 
masses are a few hundred GeV is favored from the muon $g-2$ for 
$\tan\beta \ltsim 10$, whereas for $\tan\beta \simeq 50$ heavier
slepton mass up to 
$\sim$ 1000 GeV can account for the reported
3.2 $\sigma$ difference between its experimental value and 
the Standard Model (SM) prediction.
As for the electroweak measurements, the SM gives a good description, 
and the sfermions lighter than 200 GeV tend to make the fit worse.
We find, however, that sleptons as light as 100 to 200 GeV 
are favored also from the electroweak data, if we leave out 
the jet asymmetry data that do not agree with the leptonic
asymmetry data.  We extend the survey of the preferred 
MSSM parameters by including the constraints from the 
$b \to s \gamma$ transition,
and find favorable scenarios in the minimal
supergravity, gauge-, and mirage-mediation models of supersymmetry
breaking.
\end{abstract}

\newpage
\section{Introduction}
Despite anticipation that physics beyond the Standard Model 
(SM) should show up at energies just above the reach of the 
present collider experiments, we have so far been unsuccessful 
in identifying the nature of new physics. 
Supersymmetric (SUSY) extensions of the SM, in particular,
the minimal supersymmetric 
standard model (MSSM) has several attractive features,
such as the unification of the three gauge couplings.
In Ref.~\cite{CH2000ew}, two of us performed
a comprehensive study of constraints on the MSSM 
parameters from precision electroweak (EW) experiments by using 
the data published in 1999~\cite{1998lep}. 
It was found that almost all SUSY particle masses are constrained 
to be larger than a few 100 GeV, except for the light 
wino-like chargino whose contribution improved the SM fit 
slightly. 
%%%---- new paragraph 

%%%---- new paragraph 
Since then, there have been several 
improvements in the EW measurements
and theoretical analyses.
Most notably, the LEP results have been finalized~\cite{:2005em, LEPEWWG} 
and the estimate of the running QED coupling constant at 
the $Z$ boson mass scale has been improved by the 
contribution from the BES experiment~\cite{BES, BES09}. 
Also, the measurements of the muon anomalous magnetic
moment at BNL have been finalized~\cite{Bennett:2006fi}, and
a 3.2 $\sigma$ discrepancy from the SM prediction has been 
reported~\cite{TTtalkTau2010} 
(see Refs.~[9--11]
for the determination of the hadronic contribution to the muon 
$g-2$ from other groups).
Those changes in the EW data, the $W$ boson mass 
data~\cite{m_W-Refs, PDG10} and the top-quark mass
data~\cite{PDG10, m_t-Refs} lead to 
new constraints on the MSSM parameters. 
On theory side, a number of groups
have studied the EW fits in the SM and in the MSSM:
for reviews, see e.\ g.\ Ref.~\cite{SUSYEWreviews}.
Refs.~[16--18]
update the constraints on 
the Higgs boson mass in the SM using the EW data
as well as the LEP and the Tevatron data for direct searches
for the SM Higgs boson.
Ref.~\cite{Heinemeyer:2007bw} provides
a comprehensive analysis of the MSSM EW precision fits using
the state-of-the-art multi-loop calculations of the EW
precision observables~\cite{MSSMEWPOmultiloop}. 
The EW precision fits in the minimal supergravity
mediated SUSY breaking (mSUGRA) scenarios have been
studied in many papers~[21--47].
They are also studied in the split SUSY 
model~\cite{Marandella:2005wc, Martin:2004id},
in the gauge-~\cite{Bechtle:2009ty, AbdusSalam:2009tr,
Heinemeyer:2008fb, Marandella:2005wc, Erler:1998ur, Fuks:2008ab},
anomaly-~\cite{AbdusSalam:2009tr, Heinemeyer:2008fb, Marandella:2005wc},
moduli-~\cite{AbdusSalam:2009tr, Allanach:2008tu},
and radion mediations~\cite{Marandella:2005wc},
in the supergravity models with non-universal gaugino 
masses~\cite{Belanger:2004ag}
and with non-universal Higgs masses~\cite{Buchmueller:2009fn,
Roszkowski:2009sm} and in the 25-parameter ``phenomenological MSSM''
model~\cite{AbdusSalam:2009qd}.
The EW observables in the MSSM are also studied 
in attempts to accommodate the NuTeV 
anomaly~\cite{Kurylov:2003zh, Kurylov:2003by}
and the discrepancy between the effective Weinberg angles
extracted from the leptonic and hadronic asymmetries~\cite{Altarelli:2001wx}.

Another advance on the theory side is a deeper understanding
of the jet asymmetries measured at the LEP experiments.
In Ref.~\cite{Hagiwara:2010cd},
QCD radiative corrections to the jet asymmetries are studied,
and it is argued that in the final report of the LEP
experiments~\cite{:2005em} the systematic errors of the jet asymmetries
might have been underestimated.  This is also a part of the motivation
to revisit the EW fits in this paper.

In this paper, we present quantitative results based on 
the muon $g-2$ and the final EW data from LEP
and SLC~\cite{:2005em, LEPEWWG}. 
In section 2, we discuss the MSSM contribution to the muon $g-2$,
and identify its preferred parameter region.
In section 3, we give our parametrization of the 
EW observables at the $Z$-pole and the mass and the width 
of the $W$ boson in the SM.
In section 4, we briefly review the MSSM contributions to the
EW precision observables.
In section 5, we explore a few SUSY breaking models and identify 
several preferred scenarios that can accommodate also the 
$b \to s \gamma$ rate.
Section 6 gives summary and discussions.

\section{The muon $g-2$ vs MSSM}

\subsection{The muon $g-2$ in the MSSM}

The muon $g-2$ is a precisely measured quantity,
and hence it is an excellent probe of new physics at the TeV scale.  
The measurement at BNL was finalized in
2006~\cite{Bennett:2006fi}.  After including
a small shift in the value of the proton-to-muon
magnetic ratio reported since then~\cite{Mohr:2008fa},
the experimental value is~\cite{PDG10}:
\begin{align}
 a_{\mu}^{\rm exp} = 11 659 208.9 (6.3) \times 10^{-10}.
\end{align}
As for the SM prediction, the recent improvement in the low-energy
$e^+e^- \to$ hadrons data from 
BaBar~\cite{Aubert:2009fg}, BES~\cite{BES09}, 
CMD-2~\cite{CMD-2}, KLOE~\cite{:2008en, Ambrosino:2010bv}
and SND~\cite{SND}
allows us to reduce the uncertainty.
In Ref.~\cite{TTtalkTau2010},
the SM prediction has been evaluated as
(after correcting a typo in Ref.~\cite{TTtalkTau2010}), 
\begin{align}
 a_{\mu}^{\rm SM} = 11 659 183.0 (5.1) \times 10^{-10},
\label{eq:a_mu:SM_HLMNT10}
\end{align}
which includes all the major updates on the $e^+e^- \to$
hadrons data, 
and adopts the estimate~\cite{Prades:2009tw},
$a_{\mu}^{\rm l{\mbox -}by{\mbox -}l} = 10.5 (2.6) \times 10^{-10},$
for the light-by-light contribution.
Other independent
analyses~\cite{Davier:2010nc, Jegerlehner:2011ti, Jegerlehner:2009ry}
based on the $e^+e^-$ data give similar estimates. 
Hence, the observed value of the muon $g-2$ is larger than
the SM prediction by 
\begin{align}
 \delta a_{\mu} \equiv
 a_{\mu}^{\rm exp} - a_{\mu}^{\rm SM} = (25.9 \pm 8.1) \times 10^{-10},
\label{eq:a_mu:the3.2sigma}
\end{align}
which differs from zero by 3.2 $\sigma$.
It is tempting to interpret the difference
as a contribution of new particles with the muon quantum number,
such as smuons in the SUSY SM.
Throughout this article we assume that the MSSM contribution
accounts for this discrepancy.

In the MSSM, the contribution to the muon $g-2$ has been calculated 
up to and including the two-loop level~\cite{Stockinger}.  In view 
of the smallness of the two-loop contribution, in this paper
we restrict our analyses in the one-loop approximation\footnote{
The effect from the $\tan\beta$-enhanced
resummation in the muon-muon-Higgs vertex can change the MSSM contribution
by about 10\%~\cite{Girrbach:2009uy},
which does not affect the conclusions in the present paper significantly.}.

\begin{figure}
\begin{center}
\includegraphics[scale=0.7]{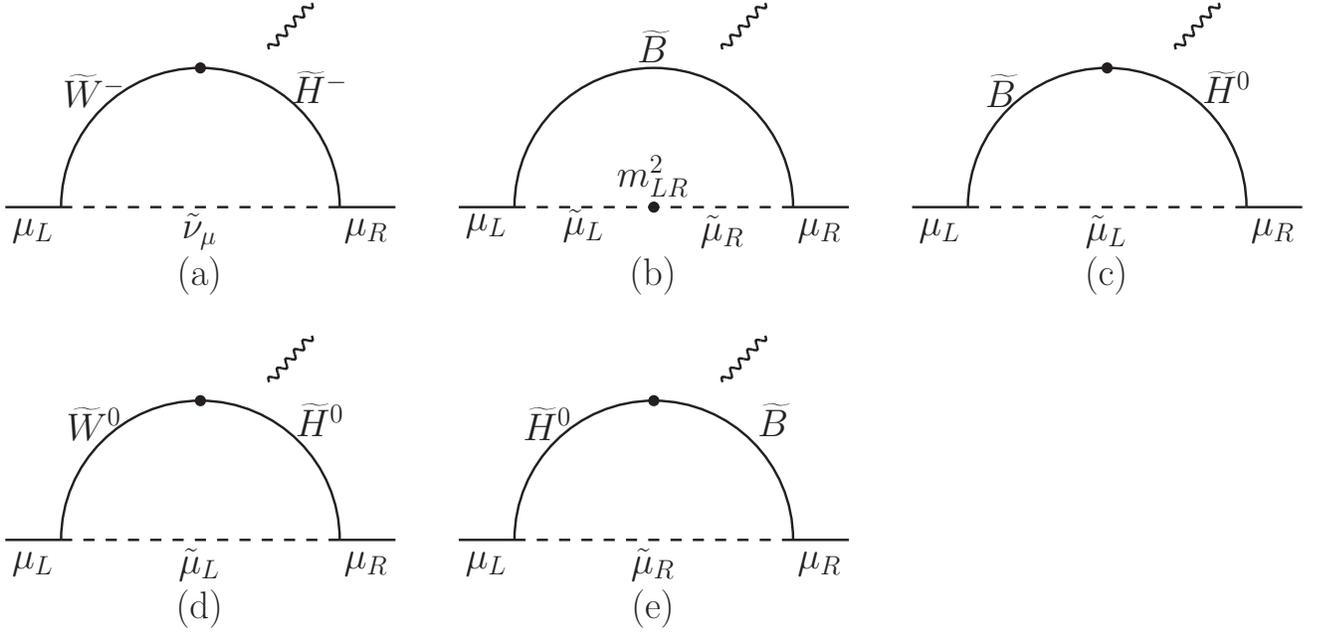}
\end{center}
\caption{The SUSY contributions to the muon $g-2$ which give the
leading terms of the expansion in $m_Z/m_{\rm SUSY}$.  The photon
(wavy line) is attached to all the charged particles.}
\label{fig:massins}
\end{figure}

At one-loop, the MSSM contribution comes from the 
chargino contribution $a_\mu(\widetilde{\chi}^-)$
and the neutralino contribution $a_\mu(\widetilde{\chi}^0)$.  
The relevant one-loop expressions in the notation of 
Ref.~\cite{CH2000ew} are found e.g.\ in Refs.~\cite{Cho:2000sf, 
Cho:2001nfa}, as:
\begin{subequations}
\begin{eqnarray}
 a_\mu(\widetilde{\chi}^-) &=&
\frac{1}{8\pi^2}\,\frac{m_\mu}{m_{\widetilde{\nu}_\mu}}
\sum_{j=1}^2
\left\{
\left(
     \left| g_L^{\widetilde{\chi}^-_j \mu \widetilde{\nu}_\mu}
          \right|^2
     +
\left| g_R^{\widetilde{\chi}^-_j \mu \widetilde{\nu}_\mu} \right|^2
\right)
\frac{m_\mu}{m_{\widetilde{\nu}_\mu}}\,
G_1 \left(
     \frac{m_{\widetilde{\chi}^-_j}^2}{m_{\widetilde{\nu}_\mu}^2}
            \right)
 \right.
\nonumber \\
 && \quad \quad \quad \quad \quad \quad
 +
 \left.
   {\rm Re}
     \left[
      \left(
       g_R^{\widetilde{\chi}^-_j \mu \widetilde{\nu}_\mu}
      \right)^*
      g_L^{\widetilde{\chi}^-_j \mu \widetilde{\nu}_\mu}
     \right]
   \frac{m_{\widetilde{\chi}^-_j}}{m_{\widetilde{\nu}_\mu}}\
   G_3 \left(
        \frac{m_{\widetilde{\chi}^-_j}^2}{m_{\widetilde{\nu}_\mu}^2}
       \right)
 \right\} ,  \\
 a_\mu(\widetilde{\chi}^0) &=&
 -
 \frac{1}{8\pi^2}
 \sum_{i=1}^2 \frac{m_\mu}{m_{\widetilde{\mu}_i}}
 \sum_{j=1}^4
 \left\{
   \left(
     \left| g_L^{\widetilde{\chi}^0_j \mu \widetilde{\mu}_i}
          \right|^2
     +
     \left| g_R^{\widetilde{\chi}^0_j \mu \widetilde{\mu}_i}
          \right|^2
   \right)
   \frac{m_\mu}{m_{\widetilde{\mu}_i}}
   G_2\left(
       \frac{m_{\widetilde{\chi}_j^0}^2}{m_{\widetilde{\mu}_i}^2}
      \right)
 \right.
\nonumber\\
 && \quad \quad \quad \quad \quad \quad \quad \quad
 \left.
  + {\rm Re}
    \left[
      \left(
       g_R^{\widetilde{\chi}^0_j \mu \widetilde{\mu}_i}
      \right)^*
      g_L^{\widetilde{\chi}^0_j \mu \widetilde{\mu}_i}
    \right]
 \frac{m_{\widetilde{\chi}_j^0}}{m_{\widetilde{\mu}_i}}\,
    G_4\left(
        \frac{m_{\widetilde{\chi}_j^0}^2}{m_{\widetilde{\mu}_i}^2}
       \right)
 \right\} , 
\end{eqnarray}
\end{subequations}
where
\begin{subequations}
\begin{align}
 G_1(x) =&  \frac{1}{12(x-1)^4}
  \left[   (x-1)(x^2 - 5x - 2) + 6 x\,{\rm ln}\,x  \right] \, , \\
 G_2(x) =&  \frac{1}{12(x-1)^4}
  \left[   (x-1)(2x^2 + 5x - 1) - 6 x^2\,{\rm ln}\,x  \right]\, , \\
 G_3(x) =&  \frac{1}{2(x-1)^3}
  \left[   (x-1)(x-3) + 2\,{\rm ln}\,x  \right]\, ,\\
 G_4(x) =&  \frac{1}{2(x-1)^3}
  \left[   (x-1)(x+1) - 2 x\,{\rm ln}\,x  \right]\, .
\end{align}
\end{subequations}
Even though these expressions are useful for numerical calculations, 
they are not particularly illuminating for the purpose of understanding their
dependences on the SUSY parameters.  The main disadvantage of
the above expressions is that they are written in terms of
the mass eigenstates, in terms of which the dependences on
the SUSY breaking parameters are hidden by the electroweak
symmetry breaking that causes complex mixings.

In the weak eigenstates, the structure of the
one-loop contributions becomes much more transparent.
This simplification occurs since
the expressions in the weak eigenstates are equivalent to the
$m_Z/m_{\rm SUSY}$ expansion, where $m_{\rm SUSY}$ is the typical
SUSY breaking mass scale.  The price we have to pay is that 
the leading terms in the expansion are 
not useful when
$m_{\rm SUSY} \sim m_Z$.  However, 
we will find below that this expansion is very useful when analyzing
the SUSY parameter dependence.

The leading terms in the $m_Z/m_{\rm SUSY}$ expansion
are given by the five diagrams (a) to (e) in Fig.~\ref{fig:massins},
whose contributions can be expressed compactly as
\begin{subequations}
\label{eq:C1-N2} 
\begin{eqnarray}
 a_\mu(\tilde{W}{\mbox -}\tilde{H}, \tilde{\nu}_\mu)
&=& 
 \phantom{-}  \frac{g^2}{8\pi^2}
\frac{m^2_\mu M_2 \mu \tan \beta}{m^4_{\tilde{\nu}}} \ 
F_a \left( \frac{M_2^2}{m^2_{\tilde{\nu}}}, 
           \frac{\mu^2}{m^2_{\tilde{\nu}}} \right), 
\label{eq:C1} \\
 a_\mu(\tilde{B}, \tilde{\mu}_L{\mbox -}\tilde{\mu}_R)
&=& 
 \phantom{-}  \frac{g_Y^{2}}{8 \pi^2}
\frac{m^2_\mu \mu \tan \beta}{M_1^3} \ 
F_b \left( \frac{m^2_{\tilde{\mu}_L}}{M_1^2}, 
           \frac{m^2_{\tilde{\mu}_R}}{M_1^2} \right),
\label{eq:N1} \\
 a_\mu(\tilde{B}{\mbox -}\tilde{H}, \tilde{\mu}_L)
&=&
\phantom{-} \frac{g_Y^{2}}{16\pi^2}
 \frac{m^2_\mu M_1 \mu \tan \beta}{m^4_{\tilde{\mu}_L}} \ 
F_b \left( \frac{M_1^2}{m^2_{\tilde{\mu}_L}}, 
           \frac{\mu^2}{m^2_{\tilde{\mu}_L}} \right), 
\label{eq:N3}  \\
 a_\mu(\tilde{W}{\mbox -}\tilde{H}, \tilde{\mu}_L)
&=& 
 - \frac{g^2}{16\pi^2}
 \frac{m^2_\mu M_2 \mu \tan \beta}{m^4_{\tilde{\mu}_L}} \
F_b \left( \frac{M_2^2}{m^2_{\tilde{\mu}_L}},
           \frac{\mu^2}{m^2_{\tilde{\mu}_L}} \right),
\label{eq:N4} \\
  a_\mu(\tilde{B}{\mbox -}\tilde{H}, \tilde{\mu}_R)
&=&
- \frac{g_Y^{2}}{8\pi^2}
 \frac{m^2_\mu M_1 \mu \tan \beta}{m^4_{\tilde{\mu}_R}} \
F_b \left( \frac{M_1^2}{m^2_{\tilde{\mu}_R}},
           \frac{\mu^2}{m^2_{\tilde{\mu}_R}} \right),
\label{eq:N2}  
\end{eqnarray}
\end{subequations}
respectively.
The functions $F_a(x,y)$ and $F_b(x,y)$ are defined as:
\begin{eqnarray}
 F_a (x, y) \equiv - \frac{G_3(x)-G_3(y)}{x-y}, ~~~~~~~~~~~
 F_b (x, y) \equiv - \frac{G_4(x)-G_4(y)}{x-y},
\end{eqnarray}
which are symmetric under exchange of the two arguments.
\begin{figure}
\begin{center}
\includegraphics[scale=0.95]{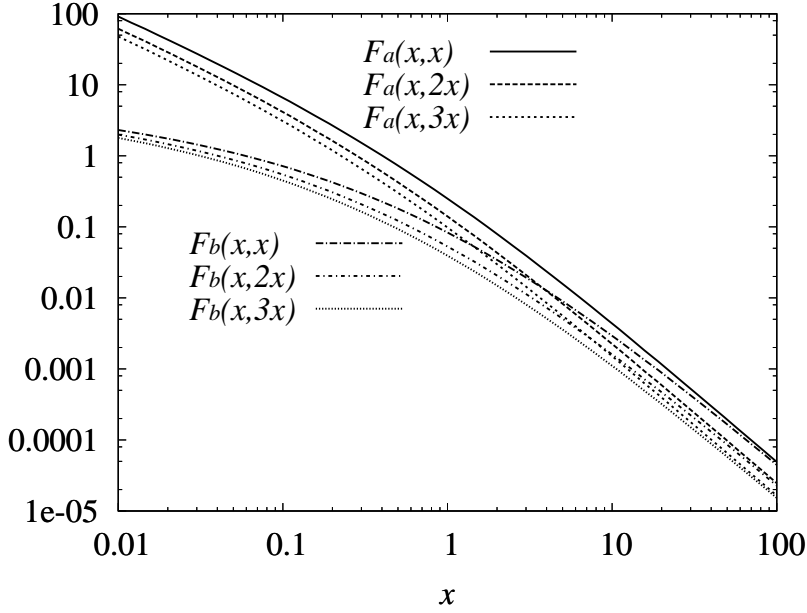}
\end{center}
\vspace*{-0.7cm}
\caption{The behaviors of the functions $F_a(x,y)$ and $F_b(x,y)$,
which appear in SUSY contributions to the muon $g-2$,
for $y=x, 2x, 3x$.}
\label{fig:Fab_funcplot}
\end{figure}
The functions $G_3(x)$ and $G_4(x)$ are monotonically decreasing 
for $0 < x < \infty$, and hence the functions $F_a(x,y)$ 
and $F_b(x,y)$ are positive for all positive $x$ and $y$.
In Fig.~\ref{fig:Fab_funcplot}, we show the behaviors of 
$F_{a,b}(x, nx) = F_{a,b}(nx, x)$ for $n=1,\ldots,3$.

The expressions (\ref{eq:C1-N2}a)-(\ref{eq:C1-N2}e)
allow us to make a few general observations on the SUSY
parameter dependences.
The first one is that
the contributions from the diagrams (a)-(c) in Fig.~\ref{fig:massins}
are positive, while those from the diagrams (d) and (e)
are negative for $M_2\mu$ and $M_1\mu>0$.
In addition, if the mass-splitting in the
$(\tilde{\nu}_\mu, \tilde{\mu}_L)$ doublet is small we can
conclude that the sum of the diagrams (a) and (d), or
that of Eqs.~(\ref{eq:C1-N2}a) and (\ref{eq:C1-N2}d) is
always positive, because $F_a(x,y)$ is always larger than $F_b(x,y)$
for the same arguments as shown in Fig.~\ref{fig:Fab_funcplot}.
If the contributions from the diagrams (b), (c), and (e) are 
suppressed, being proportional to $g_Y^2=g^2\tan^2\theta_W$,
we can conclude that $M_2\mu$ should be positive in order to 
explain the positive deviation of the muon $g-2$ from
the SM prediction in Eq.~(\ref{eq:a_mu:the3.2sigma}).

In fact, the only way to obtain a
positive MSSM contribution for $M_1\mu<0$ and $M_2\mu<0$
is to make the contribution from the diagram
(e), Eq.~(\ref{eq:C1-N2}e), dominates over the others, 
which requires $(M_2/m_{\tilde{L}}^4) 
F_a(M_2^2/m^2_{\tilde{L}}, \mu^2/m^2_{\tilde{L}}) 
\ll \tan^2\theta_W (M_1/m_{\tilde{E}}^4) 
F_b(M_1^2/m^2_{\tilde{E}}, \mu^2/m^2_{\tilde{E}})$.
This situation is realized for
$|M_1|, m_{\tilde{E}} \ll |M_2|, m_{\tilde{L}}$:
for instance, when
$(\tan\beta, \mu, m_{\tilde{E}}, m_{\tilde{L}}, M_2, M_1)=
(50, -300, 120, 1200, 1000, 100)$, where the mass dimension
is counted in units of GeV, the diagram (e) dominates
over the others, and the predicted value of $a_\mu$ is within the
1-$\sigma$ favored region. In this example, an $M_1\mu<0$ solution is
realized by setting $M_2/M_1=m_{\tilde{L}}/m_{\tilde{E}}=10$.  We find
that all the solutions with $M_1\mu<0$ require both $|M_2/M_1| \gg 1$
and $m_{\tilde{L}}/m_{\tilde{E}} \gg 1$. However, as discussed later,
scenarios with heavy left-handed sleptons do not lead to a significant
signal in the electroweak precision measurements, and we do not consider
such scenarios hereafter. 

Another interesting observation is that in the case where $\mu$
is large as compared to the SUSY breaking soft mass parameters, the
contributions from the diagrams (a) and (c)-(e) are suppressed by
$1/\mu$, as manifestly shown by
the Higgsino propagator in the diagrams Figs.~\ref{fig:massins} (a) and
(c)-(e). On the other hand, the diagram (b) is proportional to $\mu$,
which makes this contribution important.

The above observations can be made explicit by neglecting the D-term and
the F-term contributions to the slepton mass matrices, which makes
$m_{\tilde{\mu}_L} = m_{\tilde{\nu}} = m_{\tilde{L}}$ and
$m_{\tilde{\mu}_R}= m_{\tilde{E}}$, where $m_{\tilde{L}}$ and
$m_{\tilde{E}}$ are the left-handed and the right-handed slepton soft
SUSY breaking masses, respectively. Then,
Eqs.~(\ref{eq:C1})-(\ref{eq:N2}) simplify as 
\begin{subequations}
\label{eq:g-2:factorizedSUSY}
\begin{eqnarray}
 a_\mu(\tilde{W}{\mbox -}\tilde{H}, \tilde{\nu}_\mu)
&=&  \phantom{-}  a_\mu^{\rm ref} \cdot
F_a \left( \frac{M_2^2}{m^2_{\tilde{L}}}, 
           \frac{\mu^2}{m^2_{\tilde{L}}} \right),
\label{eq:g-2:factorizedSUSYa}\\
 a_\mu(\tilde{B}, \tilde{\mu}_L{\mbox -}\tilde{\mu}_R)
&=&  \phantom{-}
 a_\mu^{\rm ref} \cdot \tan^2\theta_W
\frac{m^4_{\tilde{L}}}{M_1^3 M_2}
F_b \left( \frac{m^2_{\tilde{L}}}{M_1^2}, 
           \frac{m^2_{\tilde{E}}}{M_1^2}
    \right), 
\label{eq:g-2:factorizedSUSYb} \\
 a_\mu(\tilde{B}{\mbox -}\tilde{H}, \tilde{\mu}_L)
&=& \phantom{-} a_\mu^{\rm ref} \cdot \frac12 \tan^2 \theta_W
 \frac{M_1}{M_2} \
F_b \left( \frac{M_1^2}{m^2_{\tilde{L}}}, 
           \frac{\mu^2}{m^2_{\tilde{L}}} \right),
\label{eq:g-2:factorizedSUSYc}\\
 a_\mu(\tilde{W}{\mbox -}\tilde{H}, \tilde{\mu}_L)
&=& - a_\mu^{\rm ref} \cdot \frac12 \
F_b \left( \frac{M_2^2}{m^2_{\tilde{L}}}, 
           \frac{\mu^2}{m^2_{\tilde{L}}} \right), 
\label{eq:g-2:factorizedSUSYd}\\
 a_\mu(\tilde{B}{\mbox -}\tilde{H}, \tilde{\mu}_R)
&=&
- a_\mu^{\rm ref} \cdot \tan^2\theta_W
 \frac{M_1 m^4_{\tilde{L}}}{M_2 m^4_{\tilde{E}}} \
F_b \left( \frac{M_1^2}{m^2_{\tilde{E}}},
           \frac{\mu^2}{m^2_{\tilde{E}}} \right), 
\label{eq:g-2:factorizedSUSYe}
\end{eqnarray}
\end{subequations}
where
\begin{align}
 a_\mu^{\rm ref} \equiv \frac{g^2}{8\pi^2}
                 \frac{m_\mu^2 M_2 \mu \tan\beta}{m^4_{\tilde{L}}}.
\end{align}
Only for the above specific scenario with
$m_{\tilde{L}}/m_{\tilde{E}} \gg 1$ and $|M_2/M_1| \gg 1$,
Eq.~(\ref{eq:g-2:factorizedSUSYe}) dominates over the others and
$M_1\mu<0$ is required to account for the positive shift, 
Eq.\ (\ref{eq:a_mu:the3.2sigma}).
In the very heavy Higgsino case with $|\mu/M_1|,|\mu/M_2| \gg 1$,
Eq.~(\ref{eq:g-2:factorizedSUSYb}) dominates, and $M_1\mu>0$ is
necessary to account for the positive shift.  Except for the above two
cases, Eq.~(\ref{eq:g-2:factorizedSUSYa}) dominates, and
$M_2\mu>0$ follows.

We note that the above discussion on the favored signs of $M_1\mu$ and
$M_2\mu$ are valid equally for both positive and negative $M_1$ and
$M_2$ since the signs of $M_1$ and $M_2$ enter only through the
combinations $M_1\mu$ and $M_2\mu$. %for $|A_\mu| < |\mu| \tan\beta$.
In particular, we can use the same argument to identify the favored signs
of $M_1 \mu$ and $M_2 \mu$ also for a model with 
${\rm sgn}(M_1)=-{\rm sgn}(M_2)$, which is realized in some parameter
regions of the mixed moduli and anomaly mediation model~\cite{Okumura}.

A comment on the SUSY breaking tri-linear coupling $A_\mu$ is in
order here. In the one-loop order, the parameter $A_\mu$ enters only
through the $\tilde{\mu}_L$-$\tilde{\mu}_R$ mixing in the
combination $(A_\mu^* - \mu \tan\beta)$ in the diagram
Fig.~\ref{fig:massins} (b).
In order for $A_\mu$ to affect the
muon $g-2$ significantly, Fig.~\ref{fig:massins} (b)
should
dominate over the others, but it implies large $|\mu|$ according to the
above discussions. Since there is no attractive SUSY breaking scenarios
that lead to a huge magnitude of $|A_\mu| \sim |\mu| \tan\beta$, and since
moderate values of $|A_\mu| \ltsim |\mu|$ do not affect the qualitative
behavior, we set $A_\mu=0$ in the following, unless otherwise
stated\footnote{
For the possibility of explaining the muon $g-2$
anomaly using a large $A$ term, see e.g.\ Ref.~\cite{Crivellin:2010ty}.}.

\begin{figure}
\vspace*{-1.5cm}
\hspace*{-2.cm}
\includegraphics[height=7.cm,clip]{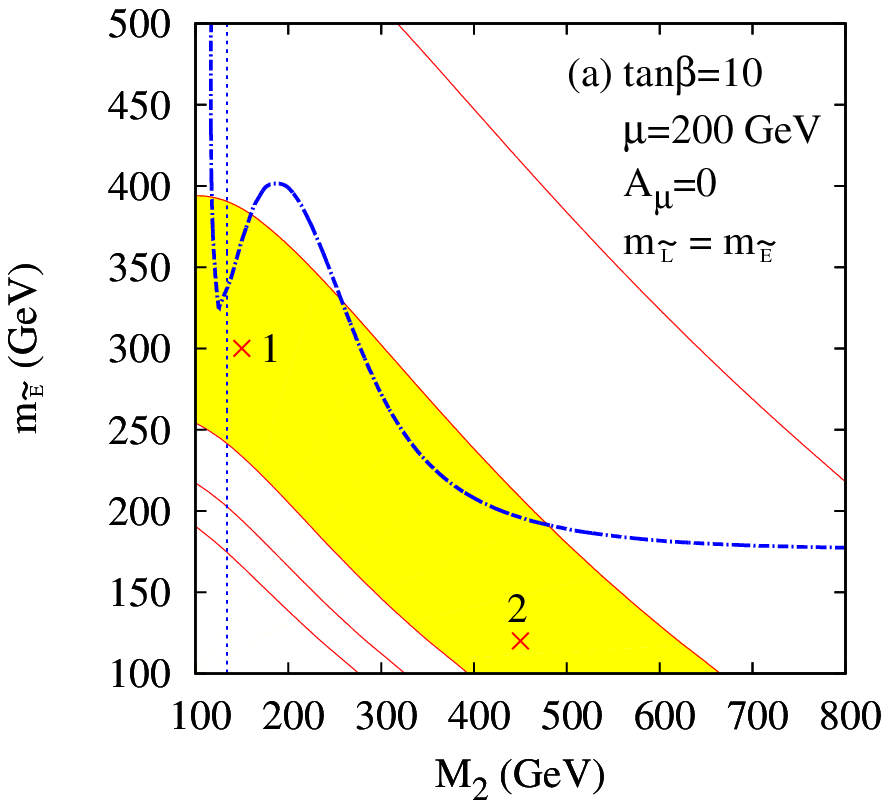}
\hspace*{-3.7cm}
\includegraphics[height=7.cm,clip]{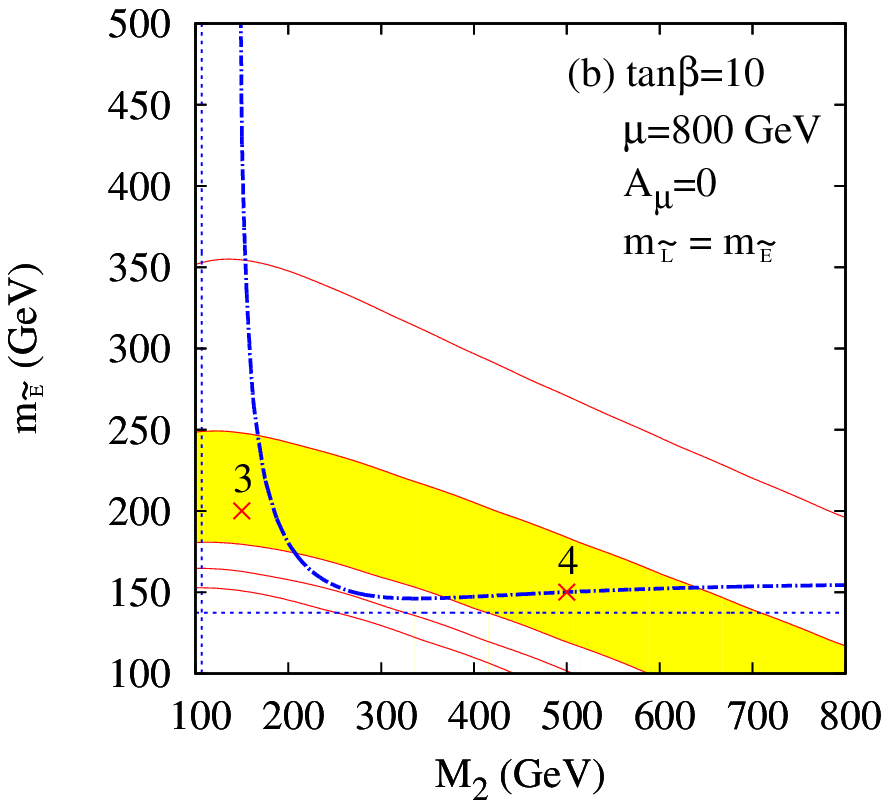}
\hspace*{-1.8cm}
\includegraphics[height=7.cm,clip]{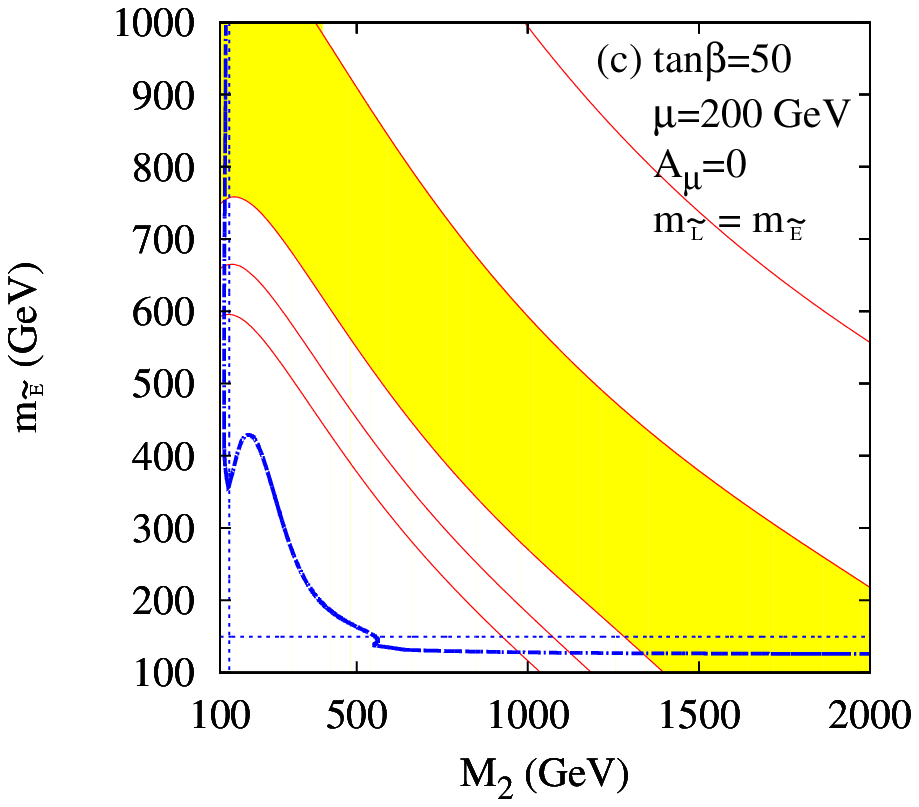}
\hspace*{-3.5cm}
\includegraphics[height=7.cm,clip]{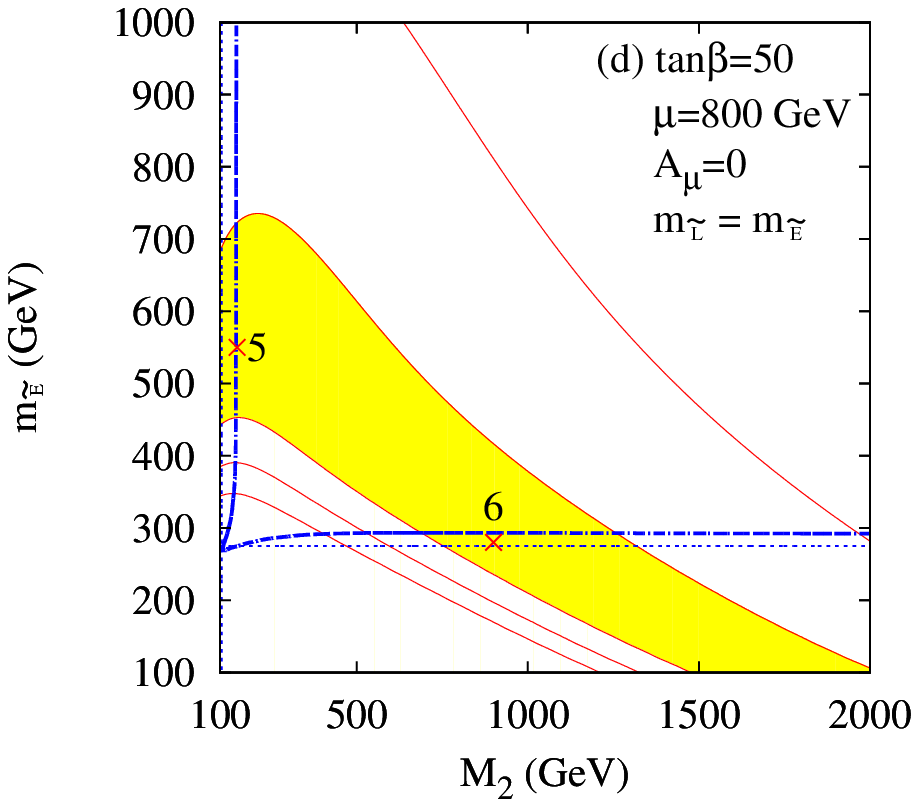}
\vspace*{-0.5cm}
\caption{The muon $g-2$, plotted against $M_2$ (the $SU(2)_L$ gaugino
 mass) and $m_{\tilde{E}}$ (the right-handed smuon soft SUSY breaking
 mass) for $\tan\beta=10$ (top two panels) and $\tan\beta=50$ (bottom
 two panels), and for $\mu=200$ GeV (left two panels) and $\mu=800$ GeV
 (right two panels). The curves are, from the lower left corner, $+3\sigma$,
 $+2\sigma$, $+1\sigma$, $-1\sigma$ and $-2\sigma$ contour
 for the difference $\delta a_\mu$ between the data and the SM
 prediction. The region on the left-hand side of the vertical dotted
 line is excluded by the chargino mass limit
 $m_{\tilde{\chi}_1^-}>103.5$ GeV~\cite{LEPSUSYWG}, and the region below the
 horizontal dotted line is ruled out by the stau mass limit
 $m_{\tilde{\tau}_1} >81.9$ GeV~\cite{PDG10}. The region below or in the
 left-hand side of the dash-dotted curve gives 
 $\Delta \chi^2_{\rm EW}>0.5$ contribution to the electroweak
 observables, see Eq.~(\ref{eq:chisqEWSUSY_SzTzMw}). The sample points
 discussed in the main text are marked by the crosses ($\times$). In the
 figures, we assume $A_\mu=0$, $m_{\tilde{L}}=m_{\tilde{E}}$ and
 $M_1/M_2 = (5/3) \tan^2 \theta_W$.} 
\label{fig:g-2}
\end{figure}

In Fig.~\ref{fig:g-2} we show contour plots of the SUSY contribution to
the muon $g-2$ as a function of the right-handed slepton mass
$m_{\tilde{E}}$ and the $SU(2)_L$ gaugino mass $M_2$. In the figures, we
examine the four combinations of $\tan\beta$ and $\mu$; $\tan\beta=10$
or 50 and $\mu=200$ or 800 GeV.
As for the other relevant SUSY parameters, we assume the
gaugino mass unification condition, $M_1/M_2 = (5/3) \tan^2\theta_W$,
and the left-handed slepton mass is taken to be equal to the
right-handed slepton mass, for simplicity. Under these
assumptions, we plot, from the lower left corner, the $+3\sigma$,  $+2\sigma$,
$+1\sigma$, $-1\sigma$ and $-2\sigma$ contours. 
(We do not plot $-3\sigma$ curves since those lie outside
of the panel for Fig.~\ref{fig:g-2}(a), (b), (c) and (d)). 
The area surrounded by the $\pm 1\sigma$ contours is shaded. The
vertical dotted lines give the lower bound on $M_2$ beyond which the
lightest chargino mass becomes less than the experimental limit, 103.5
GeV~\cite{LEPSUSYWG}. Also shown by the horizontal dotted lines are the stau
mass limit, 81.9 GeV~\cite{PDG10}. We also show the contour of a
constant ``electroweak $\chi^2$ factor'', defined as the squared sum of
the SUSY contributions to the electroweak $S_Z$- and
$T_Z$-parameters~\cite{CH2000ew} and to $m_W$;
\begin{eqnarray}
 \Delta \chi^2_{\rm EW}
= \left(\frac{(\Delta S_Z)_{\rm SUSY}}{0.04}\right)^2
+\left(\frac{(\Delta T_Z)_{\rm SUSY}}{0.04}\right)^2
+\left(\frac{(\Delta m_W)_{\rm SUSY}}{30{\rm MeV}}\right)^2 .
\label{eq:chisqEWSUSY_SzTzMw}
\end{eqnarray}
The parameter regions in the left-bottom side of
the dash-dotted curves give $\Delta\chi^2_{\rm EW}>0.5$,
and there can be a hint
of SUSY contributions to the electroweak observables in the near
future\footnote{
The ``dips'' in Figs.\ 3 (a) and (c) at $M_2\simeq 120$GeV and
350 GeV $\ltsim m_{\tilde{E}} \ltsim 400$ GeV happen since
around this region there is a destructive interference between
the combination ``$\Delta T - \ddelg/\alpha$'' and
the $\Delta R_Z$ term in $\Delta T_Z$ (see eq.\ (\ref{eq:dtz})),
which makes the contribution from $\Delta T_Z$ to
$\Delta \chi_{\rm EW}^2$ small.}. 
See discussions in Section 4 for details.

In the figures, the sample points MSSM1--MSSM6 denoted by the cross
symbols ($\times$), are chosen in the 1-$\sigma$ allowed region of the
muon $g-2$ where $\Delta \chi^2_{\rm EW}>0.5$. They represent the two
extreme cases where either a slepton (2,4,6) or an ino (1,3,5) is
light. Our choices of $\mu$ and $M_2$
allow us to cover the cases where the lighter chargino is wino-like
(3,4,5), Higgsino-like (2), or mixed (1,6). We use these six MSSM sample
points in the study of the electroweak observables in later
sections. The choice that there is either a light ino or a light slepton
is interesting in view of the electroweak study since if there is a
light new particle with nonzero electroweak quantum numbers, 
the contribution to the $S_Z$- and
$T_Z$-parameters~\cite{CH2000ew} and to $\mw$ can be significant.

For these sample points, we present separate contributions from
individual diagrams and their sum in
Table~\ref{tab:g-2:breakdownMSSM}.
In the table, we show the sum of the five
terms in the column `(a)-(e)', and the total SUSY contribution obtained
without using the $m_Z/m_{\rm SUSY}$ expansion in the column
under `total'. By comparing these two columns, we confirm for all the
six cases that the leading terms in the $m_Z/m_{\rm SUSY}$ expansion
give an excellent approximation. The last column shows the pull factor
\begin{equation}
 {\rm pull} =
  \frac{(\delta a_\mu)^{\rm SUSY}-\langle \delta a_\mu \rangle}
 {\Delta \delta a_\mu},
\label{eq:g-2pull}
\end{equation}
where the mean $\langle \delta a_\mu \rangle$ and the error
$\Delta \delta a_\mu$ are taken from Eq.~(\ref{eq:a_mu:the3.2sigma}).

For $\tan\beta=10$ and $\mu=200$ GeV shown in Fig.~\ref{fig:g-2}(a),
the 1-$\sigma$ allowed region is
very roughly given by $550{\rm GeV} \ltsim 1.7 m_{\tilde{E}} 
+M_2 \ltsim 800$ GeV. 
Here, the diagram (a) gives the dominant contribution,
as can be read off from the rows 1 and 2 in 
Table~\ref{tab:g-2:breakdownMSSM}, respectively, for the MSSM
points 1 and 2.

For a larger $\mu$ like $\mu = 800$ GeV shown in Fig.~\ref{fig:g-2}(b),
the allowed region becomes narrower, partly due to the
heavier Higgsinos, and partly because of the 
constraint from the stau mass lower bound 
$m_{\tilde{\tau}}>81.9$ GeV~\cite{PDG10}.
In the parameter region of Fig.~\ref{fig:g-2}(b), 
the diagram (b) becomes more important
because all the other diagrams are suppressed by the heavy 
Higgsino mass,
as discussed above.  This can be explicitly verified 
from Table~\ref{tab:g-2:breakdownMSSM}, in the rows 3 and 4.
In the MSSM point 3, the suppression of the diagram (a) by $1/\mu$ is not 
strong enough and hence the diagram (a) is still as important
as (b).
In the MSSM point 4, the diagram (a) is less important 
since it is suppressed not only by $1/\mu$ but also by $1/M_2$.

In the $\tan\beta=50$ case, the allowed parameter space
becomes much wider as shown in Fig.~\ref{fig:g-2}(c) 
for $\mu=200$ GeV and Fig.~\ref{fig:g-2}(d) for $\mu=800$GeV.
When $\mu \sim 200$ GeV, the favored SUSY masses
are so large that there is no region which satisfies both
the 1-$\sigma$ favored range of the muon $g-2$ and
$\Delta \chi^2_{\rm EW} > 0.5$.
When $\mu \sim$ 800 GeV, in Fig.~\ref{fig:g-2}(d), 
there appear two distinct regions of the parameters
that satisfy both conditions.
In the small $M_2$ region around the MSSM point 5,
$M_2 \gsim 100$ GeV is allowed
by the lighter chargino mass constraint, and $M_2\lsim$150 GeV gives
$\Delta \chi^2_{\rm EW} > 0.5$.
In the small $m_{\tilde{E}}$ region around the MSSM point 6, 
although $m_{\tilde{E}} \sim 300$ GeV, 
the lighter stau is as light as $\sim$100 GeV and
it gives a sizable contribution to the
$\Delta T$-parameter, which makes
$\Delta \chi^2_{\rm EW}$ non-negligible.
As for the muon $g-2$, in the MSSM point 5, 
the large slepton mass suppresses the diagram (b) despite large
$\mu$, and the light wino makes 
the diagram (a) dominate over the other contributions.
In the MSSM point 6, the large $\mu$ enhancement of the diagram (b)
is more effective, while the diagram (a) is suppressed by the large
wino mass $M_2$.  As a result, the contributions from the diagrams (a)
and (b) are comparable.

In summary, for the scenarios in which
the SUSY contribution to the muon $g-2$ can be tested by
the electroweak precision study,
the diagrams (a) and/or (b) give dominant contribution
to the muon $g-2$.

\begin{table} \begin{center} \begin{tabular}{c|cccc|ccccc|c|c|c}
\hline 
%      & $a_\mu(\tilde{W}{\mbox -}\tilde{H}, \tilde{\nu}_\mu)$
%      & $a_\mu(\tilde{B}, \tilde{\mu}_L {\mbox -}\tilde{\mu}_R)$
%      & $a_\mu(\tilde{B}{\mbox -}\tilde{H}, \tilde{\mu}_L)$
%      & $a_\mu(\tilde{W}{\mbox -}\tilde{H}, \tilde{\mu}_L)$
%      & $a_\mu(\tilde{B}{\mbox -}\tilde{H}, \tilde{\mu}_R)$
%      &  & \\
 No. & $\tan\beta$ & $\mu$ & $M_2$ & $m_{\tilde{E}}$ 
     & (a)    & (b)     & (c)  & (d)    & (e) 
     &(a)-(e) & total   & pull \\ \hline
%%%%%%%
%
1 & 10 & 200 & 150 & 300 
      & 29.6 & \hpz1.1 & 0.7 & $-2.9$ & $-1.3$
      & 27.2 & 25.0  & $-0.1$ \\
2 & 10 & 200  & 450 & 120 
      & 27.5 & \hpz8.8 & 3.3 & $-7.1$ & $-6.7$
      & 25.9 & 25.9 & $\hph 0.0$ \\
3 & 10 & 800  & 150 & 200 
      & 14.3 & 16.2 & 0.6 & $-2.7$ & $-1.3$
      & 27.1 & 27.1 & $\hph 0.1$ \\
4 & 10 & 800  & 500 & 150
      & \hpz6.9 & 21.3 & 1.0 & $-2.5$ & $-2.1$
      & 24.7& 24.3 & $-0.2$\\
5  & 50 & 800 & 150 & 550  
      & 26.9 & \hpz2.4 & 0.5 & $-2.6$ & $-1.0$
      & 26.3 & 26.0  & $\hph 0.0$\\
6 & 50 & 800 & 900 & 280
      & 18.0 & 18.0 & 2.5 & $-5.9$ & $-5.1$
      & 27.7 & 27.6  & $\hph 0.2$\\
\hline
\end{tabular}
\end{center}
\caption{The parameters for the MSSM sample points,
MSSM1 to MSSM6 in Fig.~\ref{fig:g-2}, and the breakdown of their 
contributions to the muon $g-2$ in units of $10^{-10}$.
The parameters with the mass dimension are given in GeV units,
and $A_\mu=0$ is assumed for all the points.
The numbers in the columns (a) to (e) are the contributions from
the corresponding diagrams in Fig.~\ref{fig:massins},
and the column `(a)-(e)' gives their sum.
The numbers in the column `total' are calculated 
without using the $m_Z/m_{\rm SUSY}$ expansion, which slightly 
differ from the sum of the five contributions.
The last column gives the pull factor, Eq.~(\ref{eq:g-2pull}).}
\label{tab:g-2:breakdownMSSM}
\end{table}

\subsection{The muon $g-2$ in selected SUSY breaking scenarios}
\begin{table}%[b] 
\begin{center} \begin{tabular}{l|ccccccccccc}
\hline
      & $\tan\beta$ & $\mu$   & $m_{\tilde{\mu}_L}$ & $m_{\tilde{\mu}_R}$
      & $(m_{\tilde{\tau}}^2)_{\rm LR}$
      & $A_\mu$ & $M_1$       & $M_2$ 
      & $M_3$  & $m_A$ \\ \hline
\SGone % SPS1a$^\prime$
      &  10    & $\hph$396  & 181   & 116
      & \hpz$-(88)^2$
      & $-445$ & $\hph$103  & $\hph$193
      & $\hph\hpz$572  & 425   \\
\SGtwo % SPS4$^\prime$
      &  50    & $\hph$762 & 585    & 465
      & $-(261)^2$
      & $-145$ & $\hph$277  & $\hph$510
      & $\hph$1424  & 566  \\
\GMone
      &  42    & $\hph$504 & 441   & 214
      & $-(194)^2$
      & $\hphantom{-0}$25 & $\hph$181 & $\hph$339
      & $\hph\hpz$900  & 513   \\
\GMtwo % SPS7
      &  15    & $\hph$300 & 257 &  120
      & \hpz$-(90)^2$
      & \hpz$-39$ & $\hph$169 & $\hph$327
      & $\hph\hpz$896  & 378   \\
MM1 % MAM1
      &  10    & $\hph$430  & 188  & 255
      & \hpz$-(92)^2$
      & $-465$ & $\hph$170  & $\hph$258
      & $\hph\hpz$641 &  513    \\
MM2 % MAM2
      &  10    & $-572$ & 253  & 108
      & $\hph(103)^2$
      & $\hph$245 & \hpz$-99$ & $-248$ 
      & \hpz$-847$ & 616  \\
MM3
      &  10    & $\hph534$ & 200  & 237
      & $-(102)^2$
      & $\hph$509 & $\hph$224 & $\hph$173
      & \hpz$\hph877$ & 631  \\
\hline
\end{tabular}
\end{center}
\begin{center} \begin{tabular}{l|ccc cccc}
\hline
      & $m_{\tilde{u}_L}$ & $m_{\tilde{u}_R}$ & $m_{\tilde{d}_R}$
      & $m_{\tilde{t}_L}$ & $m_{\tilde{t}_R}$ 
      & $(m^2_{\tilde{t}})_{\rm LR}$ & $(m^2_{\tilde{b}})_{\rm LR}$
       \\ \hline
\SGone % SPS1a$^\prime$
      &  \hpz526   & \hpz507 & \hpz505
      &  \hpz471   & \hpz388 
      & $-(322)^2$ & $-(153)^2$       \\ 
\SGtwo % SPS4$^\prime$
      & 1345   & 1297   & 1292
      & 1173   & 1062   
      & $-(435)^2$ & $-(435)^2$       \\
\GMone
      & 1329  & 1269 & 1263
      & 1264  & 1165 
      & $\hph(215)^2$ & $-(316)^2$      \\
\GMtwo %SPS7
      & \hpz861 & \hpz831   & \hpz829
      & \hpz836 & \hpz780  
      & $-(241)^2$ & $-(153)^2$      \\ 
MM1
      & \hpz610 & \hpz589 & \hpz546
      & \hpz556 & \hpz465 
      & $-(336)^2$ & $-(159)^2$      \\ 
MM2
      & \hpz785 & \hpz796 & \hpz823
      & \hpz689 & \hpz585 
      & $\hph(397)^2$ & $\hph(184)^2$      \\  
MM3
      & \hpz758 & \hpz731 & \hpz807
      & \hpz705 & \hpz616 
      & $-(353)^2$ & $-(183)^2$      \\  \hline
\end{tabular}
\end{center}
\caption{The values of the relevant SUSY parameters for the selected
scenarios.
The parameters with the mass dimension are given in GeV units.
$(m^2_{\tilde{f}})_{\rm LR}$ $(\tilde{f}=\tilde{\tau},
\tilde{t}, \tilde{b})$ are the left-right mixing
element in the mass-squared matrices of the sfermion $\tilde{f}$.
As for the notation of the other SUSY parameters, we use that of
Ref.~\cite{CH2000ew}.}
\label{table:sampleparam}
\end{table}

\begin{table} \begin{center} \begin{tabular}{l|ccccc|c|c|c}
\hline 
%      & $a_\mu(\tilde{W}{\mbox -}\tilde{H}, \tilde{\nu}_\mu)$
%      & $a_\mu(\tilde{B}, \tilde{\mu}_L {\mbox -}\tilde{\mu}_R)$
%      & $a_\mu(\tilde{B}{\mbox -}\tilde{H}, \tilde{\mu}_L)$
%      & $a_\mu(\tilde{W}{\mbox -}\tilde{H}, \tilde{\mu}_L)$
%      & $a_\mu(\tilde{B}{\mbox -}\tilde{H}, \tilde{\mu}_R)$
%      &  & \\
%diagram
    & (a)  & (b)     & (c)  & (d)   & (e)  &(a)-(e)& total & pull \\ \hline
\SGone % SPS1a$^\prime$
    & 25.7 & 21.5    & 1.5 & $-5.2$ & $-5.4$ & 38.1 & 37.6 & $\hph1.4$ \\
\SGtwo % SPS4$^\prime$
    & 20.0 & \hpz4.8 & 1.0 & $-3.4$ & $-2.8$ & 19.5 & 19.4 & $-0.8$ \\
\GMone
    & 34.6 & 11.7    & 1.4 & $-5.3$ & $-9.2$ & 33.2 & 33.0 & $\hph0.9$ \\
\GMtwo % SPS7
    & 27.1 & 10.6    & 1.6 & $-5.0$ & $-9.0$ & 25.3 & 24.8 & $-0.1$ \\
MM1 %MAM1
    & 19.4 & \hpz7.2 & 1.4 & $-4.5$ & $-1.9$ & 21.7 & 21.7 & $-0.5$ \\
MM2 %MAM2
    & 13.2 & 18.8    & 0.7 & $-2.7$ & $-4.2$ & 25.8 & 24.7 & $-0.1$ \\
MM3
    & 19.6 & \hpz7.9 & 1.1 & $-3.8$ & $-1.8$ & 23.0 & 23.1 & $-0.3$ \\
\hline
\end{tabular}
\end{center}
\caption{SUSY contributions to the muon $g-2$ for our sample points
in units of $10^{-10}$.  The numbers in the column `total' are calculated 
without using the $m_Z/m_{\rm SUSY}$ expansion, which slightly 
differ from the sum of the numbers in the five columns (a) to (e).
The last column gives the pull factor, Eq.~(\ref{eq:g-2pull}).}
\label{tab:g-2:breakdown}
\end{table}

In the previous subsection we have examined SUSY contributions
to the muon $g-2$ without
assuming specific SUSY breaking scenarios.
In this subsection we examine several SUSY 
breaking scenarios that are consistent with the other 
constraints like the $b \to s\gamma$ decay rate,
and discuss in detail their contributions to the muon $g-2$.
We will later examine their predictions for the electroweak observables.

We take seven scenarios that predict the muon $g-2$ values within
or very close to the 1-$\sigma$ allowed region;
a few sample points each from three SUSY breaking
scenarios, namely, the minimal supergravity (SG)~\cite{mSUGRA},
the gauge mediation (GM)~\cite{GM}, and
the mixed moduli-anomaly (MM) mediation~\cite{Okumura,Choi:2005ge} models.
We call those sample points SG1, SG2, GM1, GM2, MM1, MM2 and MM3,
respectively.

The SG1 point
is the mSUGRA sample point advocated as SPS1a$^\prime$ in 
Ref.~\cite{Aguilar-Saavedra:2005pw}, whose main advantage is
that it is compatible with all high-energy mass bounds
and with the constraints from the muon $g-2$, ${\rm Br}(b\to s\gamma)$
and the dark matter relic density.
The SG2 point is a modified version of the 
SPS4 point, which is a mSUGRA point with $\tan\beta=50$ proposed
in Ref.~\cite{Allanach:2002nj}.  
At SPS4 the unified gaugino
mass $m_{1/2}$ is 300 GeV, while at SG2 we take $m_{1/2}=650$ GeV
so that it is closer to the region favored from the muon $g-2$ and 
Br($b \to s \gamma$).  By this change in $m_{1/2}$,
the pull factors for the muon $g-2$ and ${\rm Br}(b\to s \gamma)$
are improved from 3.1 and $-5.9$ to $-0.8$ and $-1.4$, respectively.

As representatives of the gauge mediation, we take 
the GM1 and GM2 points: in GM1 $\tan\beta$ is large ($\tan\beta=42$),
while in GM2 it is moderate ($\tan\beta=15$).
At these points the lightest SUSY particle (LSP) is the
gravitino, whose
interactions are too weak to be relevant for the electroweak
observables in the present paper.
In the GM1 point, which is one of the points studied in 
Ref~\cite{Hisano-Shimizu}, 
the next-to-lightest SUSY particle (NLSP) is bino, 
while in GM2, which is suggested as SPS7 in Ref.~\cite{Allanach:2002nj},
the NLSP is the stau.
Both points fit well with the muon $g-2$ and 
Br($b\to s \gamma$).

The MM1 and MM2 points are sample points from 
the mixed moduli-anomaly (MM) mediated SUSY breaking scenario. 
In MM1, the parameter $\alpha$, which 
parametrizes the ratio between the moduli and the anomaly 
mediations, is positive, while it is negative for MM2.
In MM1 and MM2, the parameters $(l_1, l_2, l_3)$, which
parametrize the contributions from moduli to the 
gaugino masses, are taken to be $l_1=l_2=l_3=1$ so that it
allows the ``mirage unification''~\cite{Okumura},
namely the gaugino masses unify at a high scale
which can be different from the GUT scale $\sim 10^{16}$ GeV.
In the case of a positive (negative) $\alpha$,
the gaugino masses unify below (above) the GUT scale.
We take another sample point, which we call MM3, 
from a variant of the MM scenario.  At this point, 
we take $(l_1,l_2,l_3)=(1,1/2,1)$ 
so that wino is lighter than bino\footnote{
In the original KKLT model~\cite{KKLT}, 
the allowed values of $l_a$ $(a=1,\ldots,3)$ are 0 or 1. 
However, when there is a contribution from the dilaton to 
the gauge kinetic functions, it is possible to have
different predictions for the gaugino masses from the
$l_a=0$ or $l_a=1$ cases~\cite{Choi:2006xb,Abe:2005rx}.
Here we take into account such a possibility by allowing
$l_a$ to take a fractional value as an ``effective value'',
instead of explicitly introducing 
the dilaton in the gauge kinetic functions.}.  
The wino LSP is an interesting possibility
since the excess of the positron flux observed 
at PAMELA~\cite{Casolino:2008zm,Adriani:2008zr} 
can be explained by the wino dark matter~\cite{Grajek:2008pg}.

For the above seven scenarios we list in Table~\ref{table:sampleparam}
the values of the relevant SUSY parameters.  We only use the
parameters in the 
slepton and ino sectors for the study of the muon $g-2$, but
later we need the squark and the Higgs sectors for the studies of the 
EW precision observables and ${\rm Br}(b\to s\gamma)$.

The breakdown of the contributions to the muon $g-2$ at each point
with respect to the diagrams
is given in Table~\ref{tab:g-2:breakdown}.  
The discussions in the previous section 
can be verified from the numbers in this table.  For all our 
sample points, the diagram (a) gives an important contribution.  
For the points where smuons are relatively light compared to
the gauginos or the Higgsinos, such as SG1 and MM2,
the diagram (b) also gives a comparable
or larger contribution than that of the diagram (a).
For all the points, the diagrams (c)-(e) give 
only subdominant contributions.

The similarity of the SG1,\ldots,MM3 points to 
MSSM1--6 can be discussed as follows.
Since SG1 is similar to MSSM3 in the sense that 
it has a bit larger $\mu$ than the slepton and the ino masses,
both diagrams (a) and (b) give important contributions.
SG2 is similar to MSSM5 in $\mu$ and
the slepton masses but with a heavier inos, and hence the
overall size of the SUSY contribution is smaller.
GM1 can be considered to be an interpolation of MSSM5 and 6,
but with a smaller $\tan\beta$,
and hence the diagram (a) is dominant with a slightly
smaller contribution from (b).
GM2 is a relative of MSSM2, and the breakdown is similar.
GM1 and GM2 have a light right-handed
slepton and 
a moderate-mass ($\sim$200 GeV) bino, which make the contribution
from (e) more important than in other SUSY sample points.
MM1, MM2 and MM3 are similar to MSSM3, even though
they have smaller $\mu$.
At MM2, $M_1$ and $m_{\tilde{\mu}_R}$ are smaller than MM1 and MM3,
which makes the diagram (b) more important than at these two 
points.
At MM1 and MM3, $\mu$ is a bit smaller, and hence the diagram (b) 
becomes a bit less important than at MSSM3 and MM2.

In summary, similarly to the discussions
in the previous subsection, the diagrams (a) and/or (b)
give important SUSY contributions to the muon $g-2$
also in the selected SUSY breaking model points.

%%%%%%%%%%%%%%%%%%%%%%%%%%%%%%%%%%%%%%%%%%%%%%%%%%%%%%%%%%%%%%%%%
\section{The electroweak observables}
%%%%%%%%%%%%%%%%%%%%%%%%%%%%%%%%%%%%%%%%%%%%%%%%%%%%%%%%%%%%%%%%%

In this section, we briefly review the electroweak observables
in the framework of Refs.~\cite{CH2000ew,arnps98}, and update the 
parametrizations of the SM predictions.

%%%----------------------------------------------
%%% Experimental data ( 2007 Winter LEPEWWG + mt(PDG10) + alpha_s(PDG10))
%%%                   ( + m_W + Gamma_W (both from PDG10))
%%%                   ( + dalpha5had (2010HLMNT prelim.))
%%%----------------------------------------------
\def\mzdata{91.1875 (21)}     % Mz
\def\gammazdata{2.4952 (23)}  % Gamma_Z
\def\sigmahdata{41.540 (37)}  % sigma_h^0

\def\rldata{20.767 (25)}      % R_l
\def\rbdata{0.21629 (66)}     % R_b
\def\rcdata{0.1721 (30)}      % R_c

\def\altaupoldata{0.1465 (32)}  % A_l(Pole)
\def\alSLDdata{0.1513 (21)}     % A_l(SLD) = A_LR^0

\def\abdata{0.923 (20)}        % A_b
\def\acdata{0.670 (27)}        % A_c

\def\afbldata{0.01714 (95)}    % A_FB^l
\def\afbbdata{0.0992 (16)}     % A_FB^b
\def\afbcdata{0.0707 (35)}     % A_FB^c

\def\jetdata{0.2324 (12)}      % sin

\def\mwdata{80.399(23)}       % Mw (PDG10)
\def\gammawdata{2.085 (42)}    % Gam_W (PDG10)
\def\mtdata{172.0(1.6)}       % Mt (PDG10)

\def\alphasdata{0.1184(7)}     % alphas (PDG 2010)
\def\dal5hdata{0.02759(15)}   % Delta alpha_had^5 (HLMNT 2010 prelim (tau2010))

The electroweak observables of the $Z$-pole experiments are expressed in
terms of the effective $Z$ boson couplings
$g_\alpha^f$~\cite{lepewwg2000} to
$f_\alpha$, where $f$ denotes the quark/lepton species and $\alpha$
stands for their chirality. The summary of the observables in terms of
the effective couplings $g_\alpha^f$ can be found, for example, in
Refs.~\cite{CH2000ew,hhkm94}. A convenient parametrization of the
effective couplings in generic $SU(2)_L \times U(1)_Y$ electroweak
theories is given by~\cite{CH2000ew}: 
%%%------------ 
\bsub
\bea
g^{\nu}_L &=& \hphantom{-} 0.50199 + 0.45250 \dgzbarsq 
               + 0.00469 \dsbarsq + \Delta g^{\nu}_L, \\
g^{e}_L &=& -0.26920 - 0.24338 \dgzbarsq 
               + 1.00413 \dsbarsq + \Delta g^{e}_L, \\
g^{e}_R &=& \hphantom{-} 0.23207 + 0.20912 \dgzbarsq 
               + 1.00784 \dsbarsq + \Delta g^{e}_R, \\
g^{u}_L &=&  \hphantom{-} 0.34675 + 0.31309 \dgzbarsq 
               - 0.66793 \dsbarsq + \Delta g^{u}_L, \\
g^{u}_R &=& -0.15470 - 0.13942 \dgzbarsq 
               - 0.67184 \dsbarsq + \Delta g^{u}_R, \\
g^{d}_L &=& -0.42434 - 0.38279 \dgzbarsq 
               + 0.33166 \dsbarsq + \Delta g^{d}_L, \\
g^{d}_R &=&  \hphantom{-} 0.07734 +  0.06971 \dgzbarsq 
               + 0.33590 \dsbarsq + \Delta g^{d}_R, \\
g^{b}_L &=& -0.42116 - 0.38279 \dgzbarsq 
               + 0.33166 \dsbarsq + \Delta g^{b}_L, \\
g^{b}_R &=&  \hphantom{-} 0.07742 +  0.06971 \dgzbarsq 
               + 0.33590 \dsbarsq + \Delta g^{b}_R,
\label{eq:amp_sm}
\eea
\esub
%%%----------------- 
where the mean values denote the SM predictions for $m_t=172$ GeV,
$\mh=100$ GeV, $\Delta \alpha_{\rm had}^{(5)}(m_Z^2)=0.0277$
and $\hat{\alpha}_s(m_Z)_{5q}=0.118$, and the coefficients of $\dgzbarsq$ and
$\dsbarsq$ control the dependences on the oblique (gauge boson propagator)
corrections.
Here, $\dgzbarsq$ and $\dsbarsq$ are the universal gauge-boson-propagator 
corrections~\cite{hhkm94} to the effective $Z$-boson couplings and the
$Z$-$\gamma$ mixing at the $\mz$ scale, respectively, and
$\Delta g_\alpha^f$
denote the shifts due to vertex corrections. In the SM, only $(\Del
g_L^b)_{\smr}$ and $(\Del g_R^b)_{\smr}$ have non-trivial $\mt$ and 
$\mh$ dependence, and the others do not receive 
$m_t$- or $m_{H_{\rm SM}}$-dependent one-loop contribution. 
On the other hand, all the $\dgfa$ terms are non-vanishing 
at the one-loop level in the MSSM.
%%%------------

%%%------------
The universal part of the corrections, $\dgzbarsq$ and $\dsbarsq$, 
are defined as the shift in the effective couplings 
$\gzbarsq(\mzsq)$ and $\sbarsq(\mzsq)$~\cite{hhkm94}  
from their SM reference values at 
$(m_t, \mh, \Delta \alpha^{(5)}_{\rm had}(\mzsq))$ 
= (172 {\rm GeV}, 100 {\rm GeV}, 0.0277):
%%%----------------- 
\begin{subequations}
\begin{eqnarray}
\gzbarsq(\mzsq) &=& 0.55602 + \dgzbarsq ,   \\
\sbarsq(\mzsq)  &=& 0.23048 + \dsbarsq . 
\end{eqnarray}
\end{subequations}
%%%----------------- 
The shifts in the two effective couplings can conveniently be expressed 
in terms of the parameters $\dsz$, $\dtz$ and $\xa$, 
%---------------------- 
\begin{subequations}
\label{gzb_sb}
\begin{eqnarray}
\dgzbarsq &=& 0.00412 \dtz, \\
\dsbarsq  &=& 0.00360 \dsz - 0.00241 \dtz + 0.00011 \xa.
\end{eqnarray}
\end{subequations}
%---------------------- 
Here the parameter $\xa$, 
%%%-----
\begin{equation}
\xa \equiv \frac{\Delta \alpha^{(5)}_{\rm had}(\mzsq)-0.0277}{0.0003}, 
\end{equation}
%%%-----
measures the $\alpha(\mzsq)$ dependence of the effective mixing
parameter $\sbarsq(\mzsq)$. The parameters $\dtz$ and $\dsz$ denote the
shift of $S_Z$ and $T_Z$ from their values at the SM reference point,
and are related to the $S$- and $T$-parameters as~\cite{stu90}:
%
%%%-----------------------------
\begin{subequations}
\label{eq:sztzdr}
\begin{eqnarray}
\dsz &=& 
 \ds + \Delta R_Z,
\label{eq:dsz}
\\
 \dtz &=& 
   \dt -\frac{\ddelg}{\alpha} +1.49 \Delta R_Z. 
\label{eq:dtz}
\end{eqnarray}
\end{subequations}
%%%-----------------------------
The factor $\bar{\delta}_G$ is the vertex and box
corrections to the muon decay constant, $G_F$~\cite{stu90},
and $\Delta \bar{\delta}_G$ is the shift from its
SM value, $\bar{\delta}_G = 0.0055 + \Delta \bar{\delta}_G$~\cite{hhkm94}. 
The $R_Z$-parameter accounts for the difference between $S$ and 
$\sz$, and represents the running effect of the $Z$ boson propagator 
corrections between $q^2=\mzsq$ and $q^2=0$~\cite{CH2000ew}.
We define it as
\begin{equation}
 R_Z \equiv - 16 \pi \left( \frac1{\bar{g}^2_Z(m_Z^2)}  
                          - \frac1{\bar{g}^2_Z(0)}  \right),
\label{eq:def:R_Z}
\end{equation}
and $\Delta R_Z$ denotes the shift from the value of $R_Z$
at the SM reference point, 1.1879~\cite{CH2000ew}:
\begin{equation}
 R_Z = 1.1879 + \Delta R_Z.
\end{equation}

In this study, we use the $W$-boson properties, $\mw$ and $\Gamma_W$,
for the fit. Instead of 
$\Delta U$, as the third oblique parameter we take 
$\dmw = \mw -80.365({\rm GeV})$ which is given as a function of
$\ds,\dt,\du,\xa$ and $\ddelg$, as~\cite{CH2000ew}
\begin{equation}
 \Delta m_W ({\rm GeV})
 = -0.288 \Delta S + 0.418 \Delta T + 0.337 \Delta U
   -0.0055 \xa - 0.126 \frac{\Delta \bar{\delta}_G}{\alpha}.
\label{eq:Delta_m_W}
\end{equation}
%
%%%-------------
We also parametrize the $W$-boson decay width, $\Gamma_W$. To do so, it
is useful to introduce the parameter $R_W$ which parametrizes the
running of the $W$ boson coupling $\bar{g}_W(q^2)$ between the zero
momentum transfer and $q^2=m_Z^2$, since the decay width is roughly
given by 
\begin{equation}
 \Gamma_W = 3.3904  \times 10^{-1}  m_W^3  G_F
 \left( 1 + 8.478 \times 10^{-3} R_W + 0.00065 x_s \right),
\end{equation}
where, in analogy to Eq.~(\ref{eq:def:R_Z}), we define $R_W$ by
\begin{equation}
 R_W = - 16 \pi \left( \frac1{\bar{g}^2_W(m_Z^2)}  
                     - \frac1{\bar{g}^2_W(0)}  \right),
\label{eq:def:R_W}
\end{equation}
and define $\Delta R_W$ as the shift from its value at the SM reference
point:
\begin{equation}
 R_W = 2.1940 + \Delta R_W.
\end{equation}

The SM contributions to the oblique parameters, $S_Z, T_Z$,
$\mw$ and $R_Z$ are given in
Refs.~\cite{CH2000ew, HHM98} as functions of $\mt$ and $\mh$. 
We update the parametrization as
\begin{subequations}
\begin{eqnarray}
 (\Delta S_Z)_{\rm SM} &=&  0.2217 \xh -0.1188 \xh^2 +0.0320 \xh^3 
             -0.0014 \xt
             +0.0005 \xs , 
	     \label{eq:DSZ_SM} \\
 (\Delta T_Z)_{\rm SM} &=& -0.0995 \xh -0.2858 \xh^2 +0.1175 \xh^3 
                +0.0367 \xt + 0.00026 \xt^2 \nonumber \\
            & & -0.0017 \xh \xt 
                -0.0033 \xs -0.0001 \xt \xs , 
		\label{eq:DTZ_SM} \\
 (\Delta \mw)_{\rm SM} &=& 
   -0.137 \xh -0.019 \xh^2 +0.018 \xt -0.005 \xa -0.002\xs ,\\
(\Delta R_Z)_{\rm SM} &=& -0.124 \left\{
   \ln \left[ 1 + \left( \frac{26}{\mh {\rm (GeV)}} \right)^2 \right]
 - \ln \left[ 1 + \left( \frac{26}{100} \right)^2 
           \right] \right\}, \label{eq:delR_Z:parametrization}\\
 (\Delta R_W)_{\rm SM} &= &
 -0.16 \left\{
   \ln \left[ 1 + \left( \frac{23}{\mh {\rm (GeV)}} \right)^2 \right]
 - \ln \left[ 1 + \left( \frac{23}{100} \right)^2 
           \right] \right\}.  \label{eq:RW_g} 
\end{eqnarray}
\end{subequations}
The parameters $x_t$, $x_h$ and $x_s$ are defined as
\begin{eqnarray}
 x_t = \frac{m_t - 172{\rm GeV}}{3{\rm GeV}}, ~~~~~ 
 x_h = \frac{\ln(\mh/100{\rm GeV})}{\ln 10}, ~~~~~
 x_s = \frac{\hat{\alpha}_s(m_Z)_{5q} - 0.118}{0.003},
\end{eqnarray}
so that their numerical values are expected to be
less than unity.
As for the vertex corrections, the shift
$(\Delta g_{L,R}^b)_{\rm SM}$ in the SM is given by
\begin{subequations}
\begin{eqnarray}
 (\Delta g_L^b)_{\rm SM} &=& -0.000058 \xh + 0.000128 \xt  , \\
 (\Delta g_R^b)_{\rm SM} &=& -0.000042 \xh - 0.000025 \xh^4 ,
\end{eqnarray}
\end{subequations}
where the $\xh^4$ term in the equation for
$(\Delta g_R^b)_{\rm SM}$
is purely from the result of the numerical fit.

%%%%%----------------------------------------------------
%%%%%   Table: C_{fV}, C_{fA}, Delta_{Photon}^f, Delta_{EW/QCD}
%%%%%   Y.Matsumoto-kun's results
%%%%%----------------------------------------------------
\begin{table} \begin{center} \begin{tabular}{l|cccc}
   \hline
   & $C_{fV}$ & $C_{fA}$ &
     $\delta_{\Im \kappa}^f$ & $\Delta_{\rm EW/QCD}^f$ [${\rm GeV}$] \\
   \hline
   $u$     & $3.1166 +0.0030 \xs$ & $3.1377 +0.00014 \xt +0.0041 \xs$
           & 0.0000146 & $-0.000113$ \\
   $d,s$   & $3.1167 +0.0030 \xs$ & $3.0956 -0.00015 \xt +0.0019 \xs$
           & 0.0000032 & $-0.000160$ \\
   $c$     & $3.1167 +0.0030 \xs$ & $3.1369 +0.00014 \xt +0.0043 \xs$
           & 0.0000146 & $-0.000113$ \\
   $b$     & $3.1185 +0.0030 \xs$ & $3.0758 -0.00015 \xt +0.0028 \xs$
           & 0.0000026       & $-0.000040$ \\
   $\nu$   & 1 & 1            & 0       & 0 \\
   $e,\mu$ & 1 & 1            & 0.0000368 & 0\\
   $\tau$  & 1 & 0.9977       & 0.0000368 & 0\\
   \hline
\end{tabular} \end{center}
\caption{{\small
The numerical values of the factors $C_{fV}$, $C_{fA}$,
$\delta_{\Im \kappa}^f$ and $\Delta_{\rm EW/QCD}^f$
which appear in the expression for the partial widths of 
the $Z$ boson.}}
\label{tab:cvca}
\end{table}

Using the effective coupling $g_\alpha^f$, the electroweak observables
can be written in the following way. First, the $Z$-boson partial decay
width into $f\bar{f}$ is,
\begin{eqnarray}
 \Gamma_f = \frac{G_F \mz^3}{6 \sqrt{2} \pi} \biggl[ \biggl( (g_V^f)^2 +
  \delta_{\Im \kappa}^f \biggr) C_{fV} + (g_A^f)^2 C_{fA} \biggr] 
  \Biggl( 1 +
  \frac{3}{4} Q_f^2 \frac{\hat{\alpha}(\mz)}{\pi} \Biggr)
+ \Delta_{\rm EW/QCD}^f  \ .
\end{eqnarray}
The value of each correction factor is summarized in
Table~\ref{tab:cvca}. 
$C_{fV}$ and $C_{fA}$ describe the corrections to the color factor in the
vector and axial-vector currents, respectively, which have a
dependence on $\alps$ and $\mt$.
The term $\delta_{\Im \kappa}^f$ represents the corrections
from the imaginary part of loop-induced mixing of the photon
and the $Z$ boson.  The term $\Delta^f_{\rm EW/QCD}$
is the non-factorizable mixed electroweak and QCD
corrections~\cite{nonfact-EW/QCD-corr}, whose values in
Table~\ref{tab:cvca} have been copied from the second paper of
Ref.~\cite{ZFITTER}.  $Q_f$ is the electric charge of the
fermion $f$ in the normalization that $Q_f=-1$ for the electron.

\begin{figure}
\begin{center}
\includegraphics[scale=1.1]{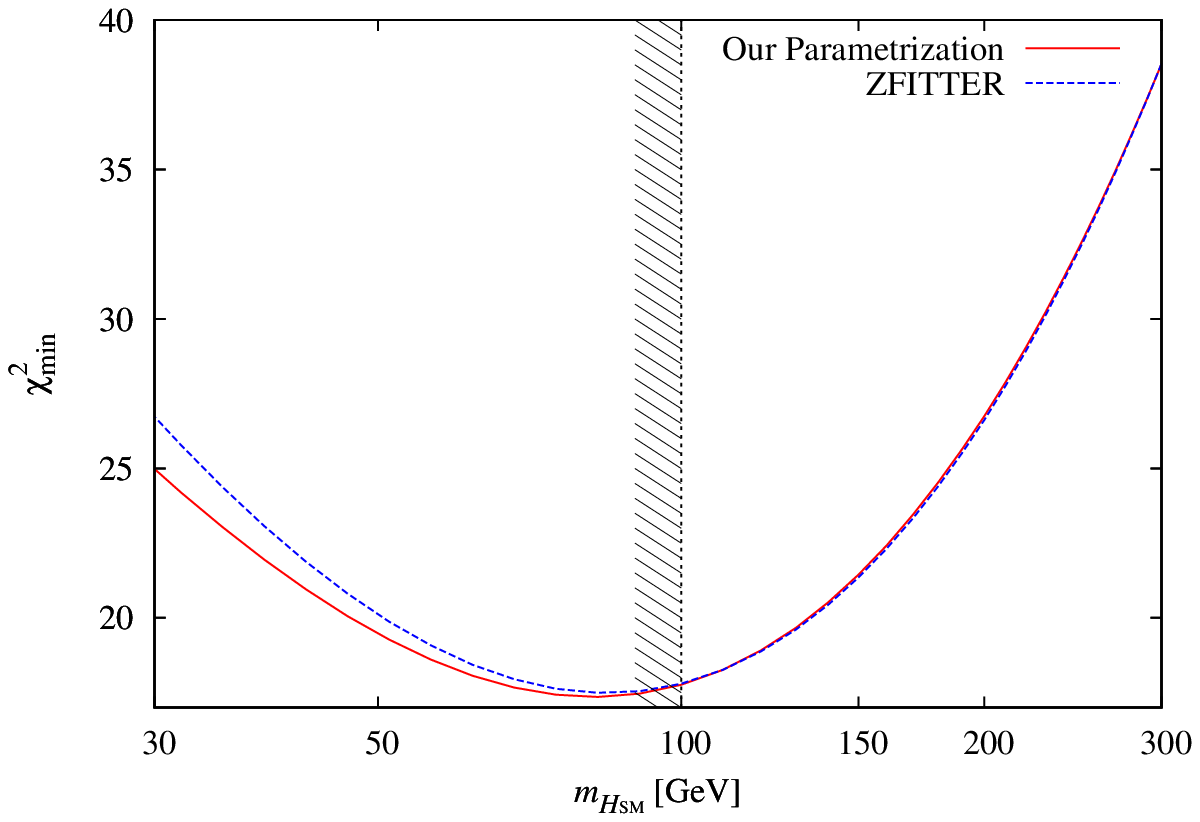}
\end{center}
\vspace*{-1.cm}
\caption{The comparison of $\chi^2_{\rm min}$ of the electroweak 
observables as a function of the SM Higgs boson mass $\mh$
fitted by using our parametrization (solid line)
and by using the output of {\tt ZFITTER} (dashed line).
Our parametrization is valid for $\mh > 100$ GeV.}
\label{fig:chisq_mh_z5}
\end{figure}

As a check of our parametrization, in Fig.~\ref{fig:chisq_mh_z5}
we give a comparison of $\chi^2_{\rm min}$ constructed from
the fit for the first 15 observables in Table~\ref{tab:ewdata2007} 
together with $m_t, \Delta \alpha_{\rm had}^{(5)}(m_Z^2), 
\hat{\alpha}_s(m_Z)_{5q}$ as a function
of $\mh$ by using {\tt ZFITTER}~\cite{ZFITTER}
and that fitted by using our parametrization.
Since our parametrization is 
designed so that it gives a good description only in 
the region $100 {\rm GeV} \le \mh \le 1000$ GeV, we find that the 
agreement becomes worse for $\mh \lsim 100$ GeV.

\begin{table}[hp]
 \begin{center}
  \rotatebox[origin=c]{0}{
  \begin{tabular}{|r|c||c|c|c|c|}
   \hline
   & data  & \multicolumn{2}{c|}{SM} & \multicolumn{2}{c|}{without $A_j$} \\ 
   \hline
   \makebox[45mm][l]{{\bf LEP 1}} & & best fit & pull & best fit & pull \\ 
   \cline{3-4} \cline{5-6}
   \begin{tabular}{l} line-shape \& FB asym.: \end{tabular} & & & & &\\
   $\Gamma_Z^{}$ (GeV) & $\gammazdata$ & 2.4958 & $-0.25$ 
                                       & 2.4963 & $-0.46$ \\
   $\sigma^0_h$(nb)    & $\sigmahdata$ & 41.478 & $\hph 1.67$ 
                                       & 41.478 & $\hph 1.69$ \\
   $~R_l^{} \; $       & $\rldata$     & 20.743 & $\hph 0.98$ 
                                       & 20.746 & $\hph 0.82$ \\
   $~A^{0,l}_{\rm FB} \; $&$\afbldata$ & 0.01647 & $\hph 0.71$
                                       & 0.01690 & $\hph 0.26$ \\
%---------
   \begin{tabular}{l}
   $\tau$ polarization: \end{tabular}
   & & & & & \\
   $A_\tau(P_{\tau})$ & $\altaupoldata$ & 0.1482 & $-0.52$ 
                                        & 0.1501 & $-1.11$ \\
%---------
   \begin{tabular}{l}
   $b$ and $c$ quark results: \end{tabular}
   & & & & & \\
   $R_b$ & $\rbdata$ & 0.21583 & $\hph 0.70$ 
                     & 0.21581 & $\hph 0.72$  \\
   $R_c$ & $\rcdata$ & 0.1722  & $-0.05$ 
                     & 0.1723  & $-0.05$ \\
   $A^{0,b}_{\rm FB}$& $\afbbdata$ & 0.1039 & $-2.92$   & ----- & ----- \\
   $A^{0,c}_{\rm FB}$& $\afbcdata$ & 0.0743 & $-1.02$   & ----- & ----- \\
%---------
   \begin{tabular}{l}
   jet charge asymmetry: \end{tabular}
   & & & & & \\
   $\sin^2\theta_{\rm eff}^{\rm lept}$ & $\jetdata$ & 
            0.2314  & $\hph 0.85$ & ----- & ----- \\
%---------
   \makebox[45mm][l]{{\bf SLC}} & & & & & \\
   $A^0_{\rm LR} (A_e)$  & $\alSLDdata$ & 
             0.1482 & $\hph 1.51$ &
             0.1501 & $\hph 0.58$ \\
   $A_b$ & $\abdata$ & 0.935 & $-0.58$      & ----- & ----- \\
   $A_c$ & $\acdata$ & 0.668 & $\hph 0.06$  & ----- & ----- \\
%---------
   \makebox[45mm][l]{{\bf Tevatron + LEP 2}}
   & & & & & \\
   $m^{}_W$ ({\rm GeV})      & $\mwdata$     & 80.376  & $\hph 0.99$ &
                                               80.400  & $-0.04$  \\
   $\Gamma^{}_W$ ({\rm GeV}) & $\gammawdata$ &  2.092 &  $-0.16$ &
                                                2.093 &  $-0.19$ \\
   \hline	
%---------
   \makebox[45mm][l]{\bf Numerical inputs } & & & & &  \\
   $m^{}_Z$ (GeV) & $\mzdata$    & ----- & ----- & ----- & ----- \\ 
   $G_F (10^{-5} {\rm GeV}^{-2}$) & $1.16637(1)$ & ----- & -----
                                                 & ----- & ----- \\
%---------
   \makebox[45mm][l]{\bf Parameters } & & & & & \\
   $\Delta \alpha_{\rm{had}}^{(5)}(m_Z^2)$  & $\dal5hdata$
                                  & 0.02761 & $- 0.14$ 
                                  & 0.02759 & $\hph 0.01$ \\ 
   $\hat{\alpha}_s(m_Z)_{5q}$&$\alphasdata$ & 0.1184 & $\hph0.00$
                                            & 0.1184 & $\hph0.06$\\ 
   $m^{}_t$ (GeV)    & $\mtdata$            & 172.3  & $-0.17$ 
                                            & 171.9  & $\hph0.05$ \\ 
   $m^{}_{H_{\rm SM}}$ (GeV) & -----        &  84.4  & -----  
                                            &  48.9  & ----- \\ 
   \hline
%---------
   $\chi^2_{\rm min}$  & & & 17.49  & & $\hph 5.35$ \\ 
   d.o.f.              & & & $18-4$ & & $13-4$ \\ 
   \hline
%---------
  \end{tabular}   }
 \end{center}
 \caption{
 The electroweak precision data,
 the SM best fit values and the pull factors.
 The SM predictions and the pull factors have been calculated 
 by using {\tt ZFITTER}~\cite{ZFITTER} by varying 
 $\mt$, $\mh$, $\hat{\alpha}_s(\mz)_{5q}$ and 
 $\Delta \alpha^{(5)}_{\rm had}(\mzsq)$ as input parameters.  
 The values of the observables are taken from Ref.~\cite{:2005em},
 except that $m_W$, $\Gamma_W$, $m_t$ and
 $\hat{\alpha}_s(m_Z)_{5q}$ are from Ref.~\cite{PDG10}, and 
 $\Delta \alpha_{\rm had}^{(5)}(m_Z^2)$ is from 
 Ref.~\cite{TTtalkTau2010}. 
 The values of $m_Z$ and  $G_F$ are fixed throughout the calculation.
 The correlation matrix elements of the $Z$ line-shape parameters 
 and those for the heavy-quark parameters 
 are found in Ref.~\cite{:2005em}.  We also show the 
 SM fit and the associated pull factors in the case where we
 do not use the jet asymmetry data, namely, $A^{0,b}_{\rm FB}$, 
 $A^{0,c}_{\rm FB}, \sin^2\theta_{\rm eff}^{\rm lept}, A_b$ and $A_c$.}
 \label{tab:ewdata2007}
\end{table}
%%%%%%%%%%%%%%%%%%%%%%%%%%%%%%%%%%%%%%%%%%%%%%%%%%%%%%%%%%%%
%%%                large table end                       %%%
%%%%%%%%%%%%%%%%%%%%%%%%%%%%%%%%%%%%%%%%%%%%%%%%%%%%%%%%%%%%

In Table~\ref{tab:ewdata2007}, we show the electroweak observables used
in the present analysis. The experimental values of the $Z$ pole
observables, including the correlations among errors that
are not reproduced in Table~\ref{tab:ewdata2007}, are taken from
Ref.~\cite{:2005em}.  The values of
$m_W, \Gamma_W, m_t$ and $\hat{\alpha}_s(m_Z)_{5q}$ are taken from
Ref.~\cite{PDG10}, and the value of $\Delta \alpha_{\rm had}^{(5)}(\mzsq)$, 
\begin{equation}
\Delta \alpha_{\rm had}^{(5)}(\mzsq) = 0.02759 \pm 0.00015,
\label{eq:Dal5h:HLMNT10}
\end{equation}
is from Ref.~\cite{TTtalkTau2010}, in which the
prediction Eq.~(\ref{eq:a_mu:SM_HLMNT10}) for the muon $g-2$ is found.
In Table~\ref{tab:ewdata2007} we also show the SM best fit values 
calculated by using {\tt ZFITTER} by varying 
$\mt$, $\mh$, $\hat{\alpha}_s(\mz)$ and 
$\Delta \alpha^{(5)}_{\rm had}(\mzsq)$
as the input parameters. $m_Z$ and  $G_F$ are fixed at their
central values throughout our analysis.

Here we consider two cases, the case using all data and the case without
using the jet asymmetry data, namely, $A^{0,b}_{FB}$, $A^{0,c}_{FB}$,
$\sin^2\theta_{\rm eff}^{\rm lept}$, $A_b$ and $A_c$, because there is
still theoretical uncertainty in the calculation of
QCD corrections~\cite{Hagiwara:2010cd}.
In the last two rows of Table~\ref{tab:ewdata2007},
we show the values of $\chi^2_{\rm min}$ and degrees-of-freedom (d.o.f.),
which is the
number of used data minus the number of input parameters.
From the fit and the value of $\chi^2_{\rm min}$, we can see
that the SM with the light Higgs boson gives a good description of the
data. If we remove the jet asymmetry data, a lighter Higgs boson is
favored. Once the best fit parameters are fixed, the corresponding
values for the 
observables can be calculated immediately, and these SM best fit values
and the associated pull factors are also shown in
Table~\ref{tab:ewdata2007}.

%\newpage

%%%%%%%%%%%%%%%%%%%%%%%%%%%%%%%%%%%%%%%%%%%%%%%%%%%%%%%%%%%%%%%%%
\section{The precision data and the MSSM}
\label{section:mssm_vs_ewdata}
%%%%%%%%%%%%%%%%%%%%%%%%%%%%%%%%%%%%%%%%%%%%%%%%%%%%%%%%%%%%%%%%%

In this section, assuming that there is new physics which 
gives rise to finite corrections to $\Delta S_Z$ and $\Delta T_Z$,
we estimate the region of $\Delta S_Z$ and $\Delta T_Z$
favored by the $Z$-pole observables.  Then, 
under the assumption that the new physics is the MSSM, we use
the constraints from $\Delta S_Z$ and $\Delta T_Z$ to find the
favored range of the MSSM parameters.  Later in this section we
use $m_W$ as another observable to constrain the favored SUSY
parameters.  We conclude this section with the discussion 
of the case where we do not use the jet asymmetry data.

\subsection{Oblique Corrections}

In this subsection, we first identify the
favored parameter range of $\Delta S_Z$ and $\Delta T_Z$.
The assumptions to compute the theoretical predictions
are the following. The input free parameters from the new
physics are taken to be $\Delta S_Z$, $\Delta T_Z$, and $\Delta g_L^b$. 
All the other vertex corrections $\Delta g_\alpha^f$ are neglected for
simplicity. As for the SM parameters, we fix $x_t$ and $x_\alpha$ at
$x_t=x_\alpha=0$ as a ``reference point''.  Consequences from different
choices of $x_t$ and $x_\alpha$ can be easily drawn, as discussed later.
We take the reference SM Higgs boson mass
to be $m_{H_{\rm SM}}=120$ GeV in this section.  As for
the other input SM parameter $x_s$, instead of fixing it at $x_s=0$, we
include it in the $\chi^2$ function, and only after finding the minimum
of the $\chi^2$ function, we integrate out $x_s$.

Using the mean values, the errors and the correlation matrix of the
observables in Ref.~\cite{:2005em}, we obtain
\begin{eqnarray}
&& \left. 
\begin{array}{l}
 \Delta S_Z = 0.020 - 2.22 \Delta g_L^b - 0.031 x_\alpha \pm 0.106 ~ \\
 \Delta T_Z = 0.053 + 0.50 \Delta g_L^b \pm 0.137 ~ \\
\end{array}
\right\} ~~ \rho_{\rm corr} = 0.91,
\label{SzTz-cont} \\
&& 
\begin{array}{l} \Delta g_L^b = -0.00033 \pm 0.00082,
\end{array} \\
&& 
\begin{array}{l} \chi^2_{\rm min} = 15.5,
\end{array}
\end{eqnarray}
where $\rho_{\rm corr}$ is the correlation between $\Delta S_Z$
and $\Delta T_Z$. 
\begin{figure}%[t]
\begin{center}
 \includegraphics[height=18.5cm,clip,bb=113 51 333 297]
                 {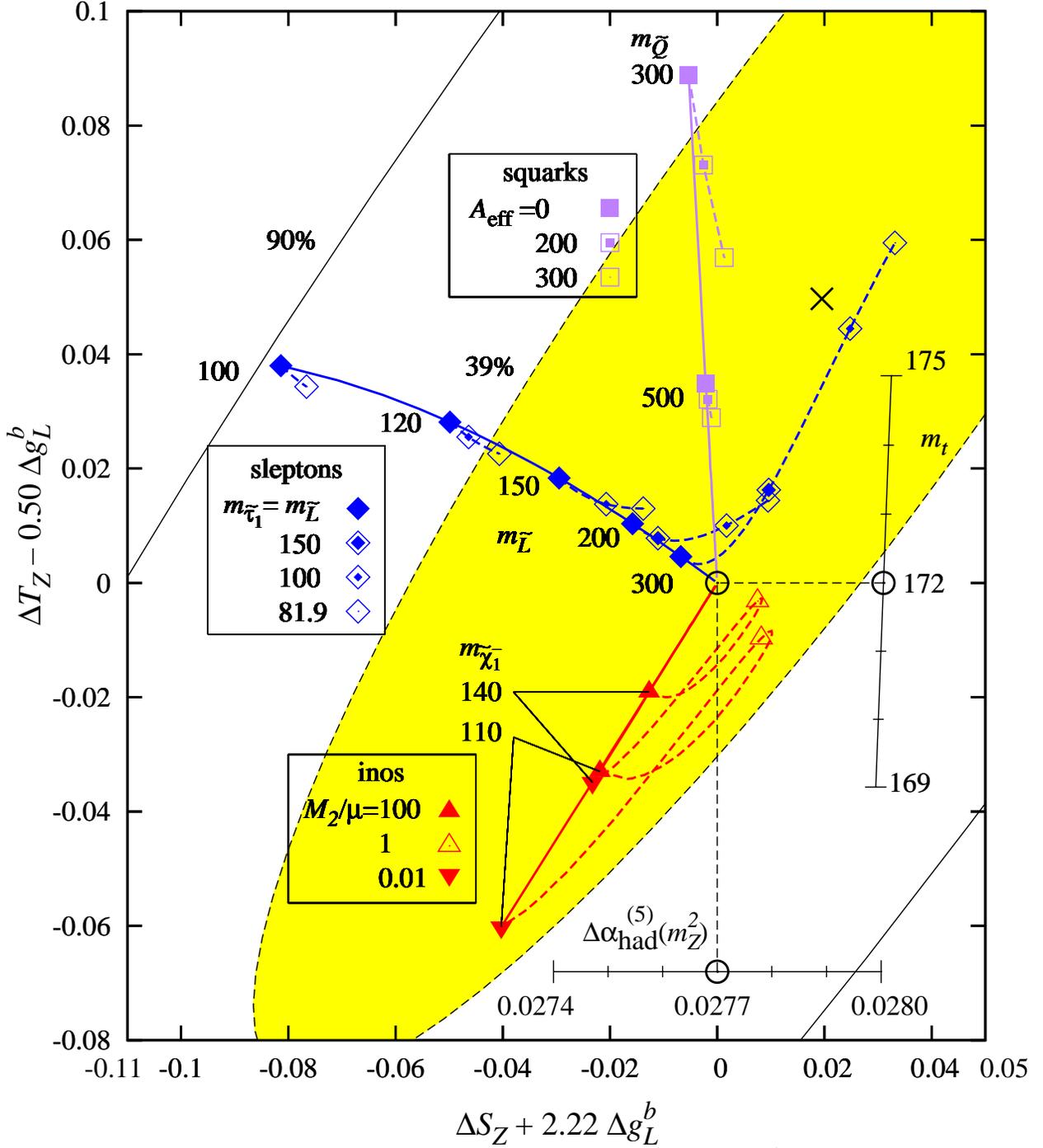} 
\end{center}
\vspace*{-1.cm}
\caption{The squark, slepton and ino contributions to 
 $(\Delta S_Z, \Delta T_Z)$ for $\tan\beta=10$. The SUSY breaking scalar
 masses for the left-handed and right-handed squarks are assumed to be
 same, denoted by $m_{\wt{Q}}$. The $\stop_L$-$\stop_R$ and
 $\sbottom_L$-$\sbottom_R$ mixings are controlled by $A_\eff = A_\eff^t
 = A_\eff^b$. The left- and right-handed sleptons are also assumed to
 have a common SUSY breaking scalar mass $m_{\wt{L}}$. The reference SM
 point, $(\mt, \mh, \Delta \alpha_{\rm had}^{(5)}(m_Z^2))=(172 {\rm
 GeV}, 120 {\rm GeV}, 0.0277)$, is marked by the open circle
 at the origin of the
 plot. If a different value of $\mt$ is chosen, then the origin would
 move according to the scale shown at the right-hand side.  Similarly,
 if a different value of $\Delta \alpha_{\rm had}^{(5)}(m_Z^2)$ is
 chosen, the origin would move according to the scale shown at the
 bottom.}
\label{fig:oblique:slsqino}
\end{figure}
In Fig.~\ref{fig:oblique:slsqino}, we show
the contours for the 39\% and the 90\% confidence levels (CL)
as shown in Eqs.~(\ref{SzTz-cont}), and also plot the SM prediction
for $\mh=120 {\rm GeV}$, $\mt=172 {\rm GeV}$ and 
$\Delta \alpha^{(5)}_{\rm had}(\mzsq)=0.0277$ as the big open circle
at the origin.
Although the $\Delta S_Z$ and $\Delta T_Z$ values which
give the minimum $\chi^2$
value are slightly different from the prediction at the
SM reference point, these shifts are
within the 1-$\sigma$ error. 
We also illustrate how the SM reference point moves according to the
change of $\mt$ from $172 {\rm GeV}$ to $175 {\rm GeV}$ and to
$\mt=169 {\rm GeV}$, as the ``ruler'' toward the right end of the figure.
As we can see, the SM prediction
for $\Delta T_Z$ becomes larger for larger $m_t$,
while $\Delta S_Z$ does not change very much
because of the stronger dependence of $\Delta T_Z$ on $\mt$, see
Eq.~(\ref{eq:DTZ_SM}). The dependence of the plot on 
$\Delta \alpha^{(5)}_{\rm had}(\mzsq)$ is shown as another ``ruler''
at the bottom of the figure.
For example, if $\Delta \alpha^{(5)}_{\rm had}(\mzsq)
- 0.0277 > 0$, then the origin moves to the right, and the 
agreement of the SM reference point to the data becomes better.

\begin{figure}%[h]
\begin{center}
\includegraphics[height=18.5cm,clip,bb=113 51 333 297]
                {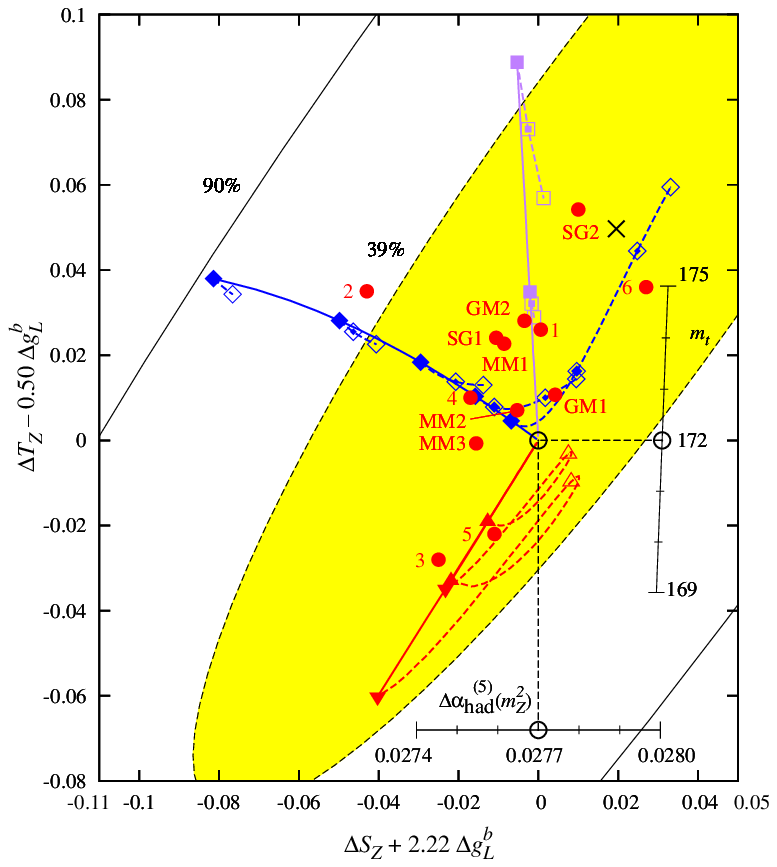} 
\end{center}
\vspace*{-0.5cm}
\caption{The predictions from our SUSY sample points, superposed
 on Fig.~\ref{fig:oblique:slsqino}.}
\label{fig:oblique:tanb10}
\end{figure}

In Fig.~\ref{fig:oblique:slsqino}, we also show separately the slepton,
squark and ino contributions to $\dsz$ and $\dtz$.
In the figure we take $\tan\beta=10$, but these SUSY contributions
do not change very much for $10 \ltsim \tan\beta \ltsim 50$.
The qualitative behaviors of those contributions 
on the $\dsz$-$\dtz$ plane have been studied in
Ref.~\cite{CH2000ew}. 

In the figure, the contributions to $\Delta S_Z$
and $\Delta T_Z$ from the sum of three generations of squarks
for the cases 
$m_{\tilde{Q}}$=300 GeV and 500 GeV without left-right mixing
among the squarks are given as the filled squares.  
In the figure, we assume $m_{\tilde{Q}} = m_{\tilde{u}_R} = 
m_{\tilde{d}_R}$ for simplicity.
The effects of the left-right mixing on these predictions
are shown by the dashed lines starting from these squares.
On each dashed line, the squark soft mass $m_{\tilde{Q}}$ is fixed at
the same value, and along
the dashed line, the parameter $A_{\rm eff}$ which controls the
left-right mixing is varied from 0 to 300 GeV.
(The definition of $A_{\rm eff}$ is the same as
in Ref.~\cite{CH2000ew}\footnote{
For completeness, the definition of $A_{\rm eff}$ is as follows:
the left-right mixing 
elements of the stop and the sbottom mass-squared matrices are given
by $m_t(A_t - \mu / \tan \beta)$ and $m_b(A_b - \mu \tan \beta)$,
respectively.  (We are neglecting possible CP-violating phases
for $\mu, A_t$ and $A_b$ for simplicity.)  We define $A_{\rm eff}^t$
and $A_{\rm eff}^b$ so that these left-right mixing elements are
equal to $m_t A_{\rm eff}^t$ and $m_b A_{\rm eff}^b$, respectively:
$A_{\rm eff}^t \equiv A_t - \mu / \tan\beta, A_{\rm eff}^b
\equiv A_b - \mu \tan\beta$. In this paper, for simplicity
we only consider the case $A_{\rm eff}^t = A_{\rm eff}^b$,
and denote this common value by $A_{\rm eff}$.}.)
The predictions for 
$A_{\rm eff}$ = 200 GeV and 300 GeV are shown by the
different squares on the dashed lines.

Similarly, the contributions to $\Delta S_Z$ and
$\Delta T_Z$  from the sum of three generations of sleptons
in the cases without left-right mixing are given by the filled
diamonds labeled as $m_{\tilde{L}}$ = 100, 120, $\ldots$, 300 GeV.
In the figure, for simplicity, we assume
$m_{\tilde{L}}=m_{\tilde{E}}$,
similarly to the squark case.
Attached to these diamonds are the dashed lines which show
the effects of the left-right mixing on these predictions.
On each dashed line, the slepton soft mass $m_{\tilde{L}}$ is fixed
at the same value, and the size of the left-right mixing
is varied by using the lighter stau mass as the measure of
the left-right mixing.  In the figure, the cases where
the lighter stau masses are 150, 100 and 81.9 GeV are shown by
the different diamonds.

In Fig.\ \ref{fig:oblique:slsqino}, the ino contributions are
also shown.  The filled upward triangles are the predictions for the cases
where the lighter chargino masses are 110 GeV and 140 GeV,
respectively, in the Higgsino-like chargino cases
with the ratio $M_2/\mu$ fixed at 100.  On each dashed
line the lighter chargino mass is fixed at the same
value, while the ratio $M_2/\mu$ is varied from 100 to 0.01
along the dashed line.

The contributions to $\Delta \sz$ from the squarks and sleptons
can be understood as follows~\cite{CH2000ew}. 
The $\Delta \sz$-parameter is defined as the sum of 
$\Delta S$ and $\Delta R_Z$.  When the left-right mixings of
the sfermions are negligible, to one-loop order,
$\Delta S$ receives contributions from left-handed
sfermions, and is proportional to the hypercharge $Y_f$ of the
sfermion $\tilde{f}$ in the loop.
The sign of the hypercharge is opposite
between the left-handed squarks ($Y_{q_L}=+1/6$) and the left-handed
sleptons ($Y_{\ell_L}=-1/2$), and
this determines the sign of $\Delta \sz$ in the limit of
no left-right mixing.  On the other hand, the sign of
$\Delta R_Z$-parameter is always negative for both squarks and sleptons
contributions~\cite{CH2000ew}, and it adds up with $\Delta S$
constructively
for sleptons while destructively for squarks. This is why
$\dsz$ is negative for the sleptons and almost zero for the squarks.

The $\Delta T_Z$-parameter is also defined as a linear
combination of $\Delta T$ and $\Delta R_Z$
with small corrections from $\overline{\delta}_G - 0.0055$.
As mentioned above, $\Delta R_Z$ is negative, but
its magnitude for the sfermions
is tiny compared to the contribution to $\Delta T$~\cite{CH2000ew}. 

To discuss the contributions
to the $\Delta T$-parameter from the sfermion sector,
it is useful to separate three cases depending on the
size of the left-right mixing of the sfermion: cases without
the left-right mixing, with small left-right mixing and
with large left-right mixing.
First, in the case without left-right mixing,
the contributions from the third generation squarks
can be written as
\begin{align}
 (\Delta T)_{\tilde{t}_L \mbox{-} \tilde{b}_L}
=& \frac{G_F C_q}{12 \sqrt{2} \pi^2 \alpha}
\frac{(m^2_{\tilde{t}_L}-m^2_{\tilde{b}_L})^2}
{m^2_{\tilde{t}_L} + m^2_{\tilde{b}_L}}
\left[ 1 + {\cal O} \left( \left(
  \frac{(m^2_{\tilde{t}_L}-m^2_{\tilde{b}_L})^2}
       {m^2_{\tilde{t}_L} + m^2_{\tilde{b}_L}}
 \right)^2 \right)
\right],
\end{align}
where $C_q$ is the color factor $(C_q=3$ for the squarks) 
and we take the limit where the squarks are
heavy compared to $m_t$.  The slepton contribution can be
obtained by the obvious replacements $C_q \to 1$, 
$\tilde{t}_L \to \tilde{\nu}$ and
$\tilde{b}_L \to \tilde{e}_L$.
Second, when left-right mixing is small enough, 
the $T$-parameter decreases as $A_{\rm eff}$ increases,
as studied in Ref.~\cite{CH2000ew}.
Third, when in the limit that left-right mixing is large,
the $T$-parameter increases as $A_{\rm eff}$ 
increases~\cite{dreeshagiwara}.
The behavior of the stau contribution to $\Delta T_Z$
interpolates the above two limits.

The ino contributions are small in general once we impose the
experimental constraint from the direct searches on the lightest
chargino mass, unless the ino masses are close to the experimental
bounds.  When inos are light,
the contributions to $\Delta R_Z$ can be sizable,
which make negative contributions to $S_Z$ and $T_Z$~\cite{CH2000ew}.  
In Fig.~\ref{fig:oblique:slsqino}, we show the 
cases $m_{\wt{\chi}^-_1} \ge 110 {\rm GeV}$
for $M_2/\mu=0.01, 1$ and 100.  For the cases
$M_2/\mu=0.01$ and 100, the predicted trajectories for
$(\Delta S_Z, \Delta T_Z)$ overlap on the line
$\Delta T_Z = 1.49 \Delta S_Z$.  This can be understood
in the following way.  Since the
wino mass parameter $M_2$ and the Higgsino mass parameter $\mu$
do not break the $SU(2)_L\times U(1)_Y$ symmetry, 
the contribution from the ino sector to the $S$- and $T$-parameters
can only come from the off-diagonal elements of the ino mass matrices.
When $M_2/\mu=0.01$ or 100, the mixing of the ino mass matrices
are suppressed by $m_Z^2/(M_2^2-\mu^2)$, which is small once
we impose $m_{\wt{\chi}^-_1} \ge 110 {\rm GeV}$ and 
the strong hierarchy between $M_2$ and $\mu$.
Hence for these hierarchical cases, the
contributions to $\Delta S_Z$ and $\Delta T_Z$ only come from 
$\Delta R_Z$, namely, $\Delta S_Z = \Delta R_Z$ and
$\Delta T_Z=1.49 \Delta R_Z$, which makes the trajectories
on Fig.~\ref{fig:oblique:slsqino} overlap.

In Fig.~\ref{fig:oblique:tanb10}, we plot our SUSY sample points on the
same frame as that of Fig.~\ref{fig:oblique:slsqino}.

In the cases of the six MSSM sample points, we can ignore the squark
contributions because we set all the squark masses to be 2 TeV. 
So, almost all the MSSM
scenarios are put near the slepton lines or ino lines. Only the sample
point 1 is apart from both lines, because it has a sizable contribution from
$\Delta \bar{\delta}_G$ to $\Delta T_Z$ by 0.036.
Although it has the slepton contribution at
$m_{\tilde{E}}=300 {\rm GeV}$ with small mass splitting and the ino
contribution at $M_2/\mu=0.75$ and $m_{\wt{\chi}_1^-}=115 {\rm GeV}$, these
contribution are almost canceled out.
The sample point 2 has
a quite large contribution from slepton, because it has light sleptons.
And it shifts
above the solid line which shows the slepton contribution without
the left-right mixing case
due to the $\Delta \bar{\delta}_G$ contribution
to $\Delta T_Z$ by 0.015.   The sample points 3 and 5
are characteristic in the ino contribution,
because of a light chargino mass, 
$m_{\wt{\chi}_1^-}\sim150 {\rm GeV}$, and a small
$M_2/\mu$ ratio, $M_2/\mu=0.19$.  In particular, the sample point 5 has
essentially only ino contribution because of heavy sleptons, 
$\sim 800 {\rm GeV}$.  On the other hand,
the sample point 3 has a non-negligible slepton contribution 
($m_{\tilde{E}}\sim 200 {\rm GeV}$) compared to the sample point 5, and
a small contribution from $\Delta \bar{\delta}_G$ of $-0.010$ to
$\Delta T_Z$.
The sample point 4 is determined almost only by slepton contribution 
from
$m_{\tilde{E}} =150 {\rm GeV}$ and $m_{\tilde{\tau}_1}\sim100 {\rm GeV}$.
The sample point 6
is an interesting point, because it is the case of large SUSY breaking mass,
$m_{\tilde{E}}=300 {\rm GeV}$ with a light stau,
$m_{\tilde{\tau}_1}=144 {\rm GeV}$.
In this parameter region $\dt$ increase as $A_\eff$ or
squared mass difference increase.

As for the predictions from the selected scenario points,
for all the selected model points except MM3,
the ino contributions are negligible 
because there $m_{\wt{\chi}_1^-}\ge200 {\rm GeV}$.
The SG2 scenario gives
the largest contribution to $\dtz$. This contribution mainly comes from
slepton with a large $m_{\tilde{E}}$ with a large squared mass
splitting,
and the contributions from the other sectors are negligible because
the squark masses are more than
1 TeV.  The GM2 and MM2 scenarios also have heavy squarks, and in the large
SUSY breaking mass with light stau region, 
$(m_{\tilde{E}},m_{\tilde{\tau}_1})=(250 {\rm GeV},120 {\rm GeV})$ and
$(245 {\rm GeV},104 {\rm GeV})$, respectively. 
The SG1 and MM1 scenarios have similar
parameters, and are
located at almost the same place in the $\dsz$-$\dtz$
plane. At the GM1 point, we can neglect the slepton contribution
because of large mass, $\sim 440 {\rm GeV}$, and it has the contribution from
squarks at $m_{\tilde{q}}\sim800 {\rm GeV}$ and the 
$\Delta \bar{\delta}_G$ contribution, 0.014. 

In Figs.~\ref{fig:oblique:slsqino} and \ref{fig:oblique:tanb10}, 
we do not show the contributions from Higgs bosons, which
are known to be 
small for $m_h > 115$ GeV~\cite{CH2000ew}.  In fact, for
our selected scenarios SG1,\ldots,MM3, the
contributions from the Higgs sector is 
$0<\Delta S_Z \lsim 0.004$ and $-0.003 \lsim \Delta T_Z<0$.

\subsection{Data without Jet Asymmetry and Oblique Corrections}

It may be worth repeating the above analysis
after excluding the jet asymmetry data from the input $Z$-pole
precision observables.

\begin{figure}
\begin{center}
\includegraphics{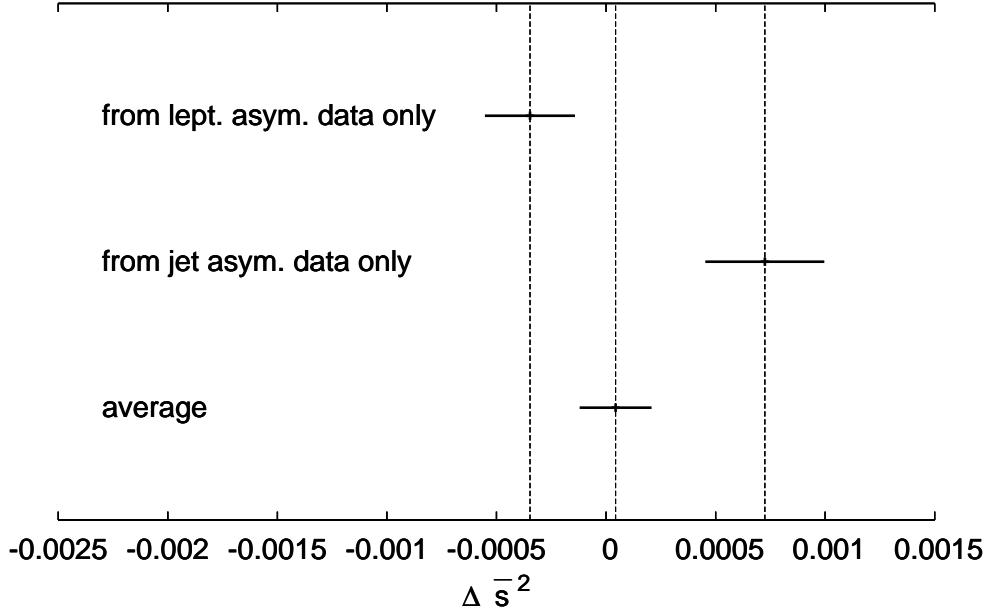}
\end{center}
\vspace*{-1.cm}
\caption{The favored ranges of $\Delta \bar{s}^2$ 
from the leptonic and the jet asymmetry data.
Also shown is the favored value for the case 
where we combine all the leptonic and the jet asymmetry data.}
\label{fig:dsbarsq}
\end{figure}

The pull factor in Table~\ref{tab:ewdata2007} shows that the data of
$b$-jet forward-backward asymmetry differ from the theoretical
expectation by roughly three standard-deviations.  
This can be seen more clearly if we look into the
favored region of $\Delta \bar{s}^2$ separately from leptonic
asymmetry and jet asymmetry.
The results are summarized in Fig.~\ref{fig:dsbarsq}. 
From the figure we can see 
that the value of $\Delta \bar{s}^2$ determined from 
the leptonic asymmetry data 
($A_{\rm FB}^{0,l}$, $A_\tau(P_\tau)$, $A_{\rm LR}^0$)
does not agree very well with
that determined from the jet asymmetry data
($A_{\rm FB}^{0,b}$, $A_{\rm FB}^{0,c}$, 
$\sin^2 \theta^{\rm lept}_{\rm eff}$, $A_b$, $A_c$).
In fact, the values for $\Delta \bar{s}^2$
for these cases are separately
\begin{eqnarray}
 \Delta \bar{s}^2 &=& -0.00035 \pm 0.00021 {\rm ~~~(lepton~only)}, 
\label{eq:leptasymsbarsq} \\
 \Delta \bar{s}^2 &=& +0.00072 \pm 0.00027 {\rm ~~~(jet~only)} .
\label{eq:jetasymsbarsq}
\end{eqnarray}
The fit for the lepton asymmetry data gives 
 $\chi^2_{\rm min}/{\rm d.o.f.}=1.64/(3-1)$, or the probability
$44\%$.  On the other hand, the fit for the jet asymmetry gives 
$0.39/(5-1)$, or the probability $98\%$. 
The values in Eqs.~(\ref{eq:leptasymsbarsq})
and (\ref{eq:jetasymsbarsq}) differ by 3.1 $\sigma$.
If we average the two values blindly, we obtain
\begin{eqnarray}
 \Delta \bar{s}^2 = 0.000044 \pm 0.00016 {\rm ~~~(all~asymmetry~data)} ,
\label{eq:allasym_sbarsq}
\end{eqnarray}
with $\chi^2_{\rm min}=9.9$.
This implies that the asymmetry data agree well
within the leptonic data and the jet data separately, but
not very well between the two sets.
Although the same result (\ref{eq:allasym_sbarsq}) is obtained
by averaging all the asymmetry data at once, with 
$\chi^2_{\rm min}/{\rm d.o.f.}=11.9/(8-1)$, we feel that this low
value of $\chi^2_{\rm min}/{\rm d.o.f.}$ is an artifact caused
by using data with large statistical errors.  Since we take seriously
the possible deviation from the SM in the muon $g-2$, 
we would like to take the difference between (\ref{eq:leptasymsbarsq})
and (\ref{eq:jetasymsbarsq}) seriously.

Recently, the jet angular distribution in $e^+e^-$ annihilation
has been re-examined~\cite{Hagiwara:2010cd} in the framework
of soft-collinear effective theory~\cite{SCET}
and a local current-three-parton ($q\bar{q}g$) operator which
contributes to the reduction of the forward-backward asymmetry
has been identified, and the associated parton shower (jet
function) has been obtained in the NLL approximation of massless
QCD.  Although the quantitative effect estimated
in Ref.~\cite{Hagiwara:2010cd} reduces
the discrepancy between the quark and lepton measurements only
slightly, the observation suggests that we may need to develop
a parton shower program which is capable of simulating the jet
angular distribution with the accuracy matching that of the
precision measurements.  Until the data can be re-analyzed
by using such advanced tools, it may be worthwhile to examine
consequences of dropping the constraints from all the jet
asymmetry measurements.

\begin{figure}
\begin{center}
\includegraphics[height=18.5cm,clip,bb=113 51 333 297]
                {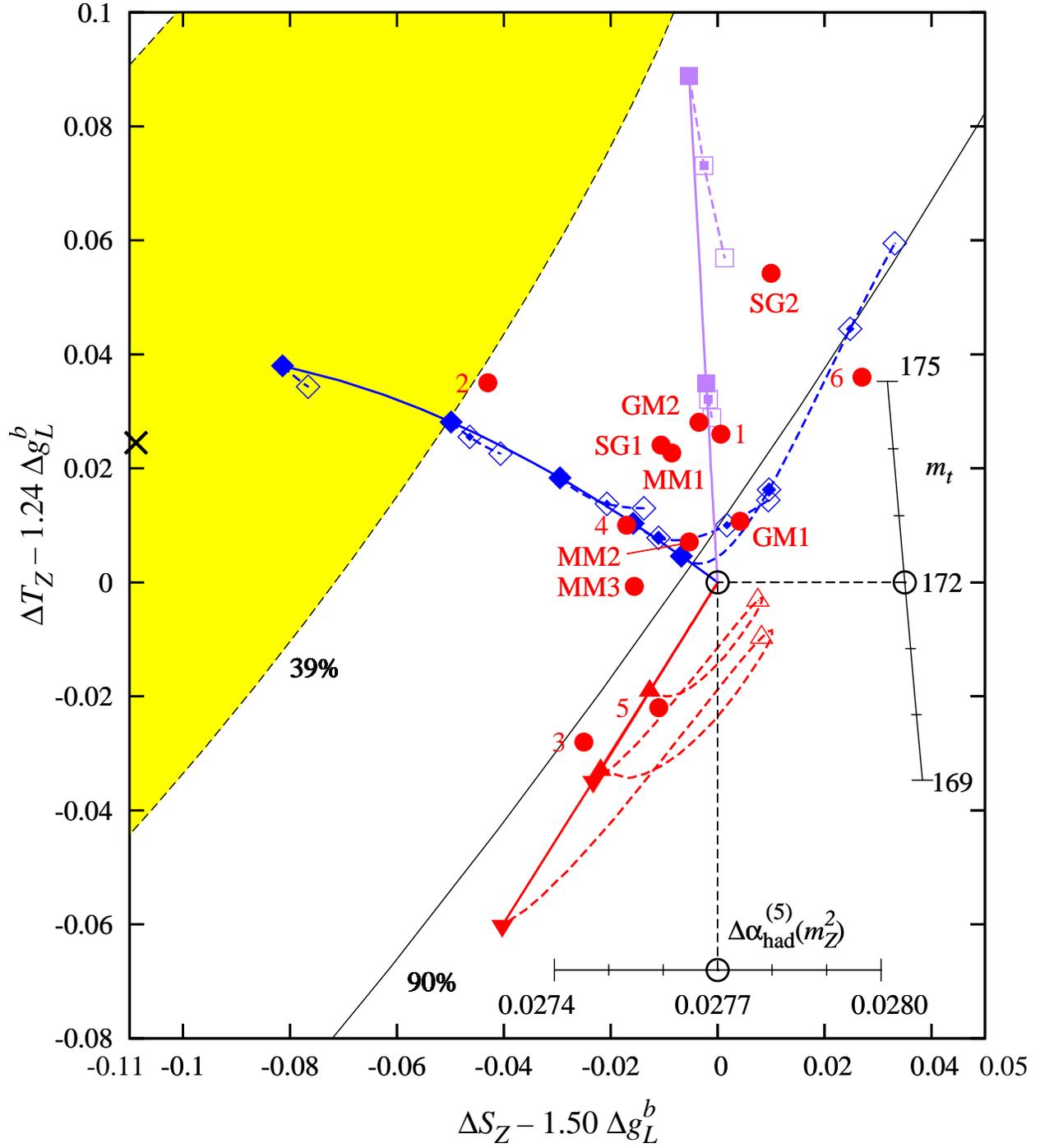}
\end{center}
\vspace*{-0.5cm}
\caption{The same figure as Fig.~\ref{fig:oblique:tanb10}
except that we omitted the jet asymmetry data.}
\label{fig:obl_nobjet}
\end{figure}

When we leave out the data for $A_{FB}^{0,b}$, $A_{FB}^{0,c}$, 
$\sin^2 \theta_{\rm eff}^{\rm lept}$, $A_b$ and $A_c$, the favored
region becomes
\begin{eqnarray}
\left.
\begin{array}{l}
 \Delta S_Z = -0.109 + 1.50 \Delta g_L^b - 0.031 \xa \pm 0.113 \\
 \Delta T_Z = \phantom{-}0.024 + 1.24 \Delta g_L^b  \pm 0.137 
\end{array}
\right\} ~~ \rho_{\rm corr} = 0.87, 
\end{eqnarray}
and the value of the minimum of $\chi^2$ is
\begin{eqnarray}
 \chi^2_{\rm min} &=& 4.7 +
 \left( \frac{\Delta g_L^b + 0.00058}{0.00083} \right)^2 .
\end{eqnarray}
As shown in Fig.~\ref{fig:obl_nobjet}, we find that the favored
region has been shifted to the negative $\Delta S_Z$ direction. We also
overlay the MSSM predictions already discussed
in Fig.~\ref{fig:oblique:tanb10}.  We find that the favored
region can be reached by relatively light ($\sim 100-120 {\rm GeV}$)
sleptons. It is interesting to note that these light sleptons can
explain the muon $g-2$ anomaly naturally.

%---------------------------------------------------
\subsection{$\mw$ in the MSSM}
%---------------------------------------------------
\begin{figure}
\begin{center}
\includegraphics[height=18.5cm,clip,bb=133 51 323 296]{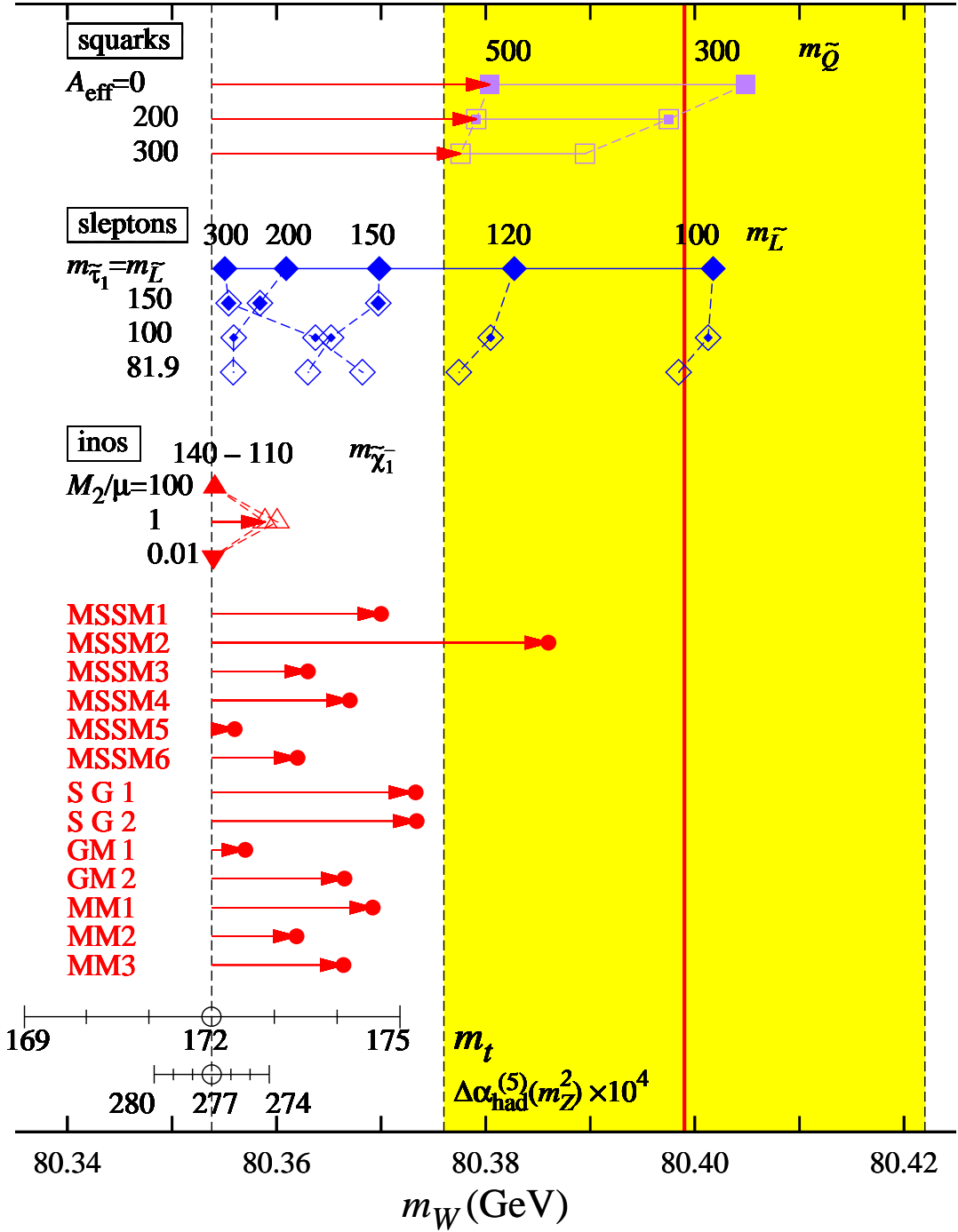}
\end{center}
\vspace*{-0.5cm}
\caption{The MSSM predictions for the $W$ boson mass from the
 squark, slepton and ino sectors, and from the MSSM sample points.
 The vertical dashed line at $m_W=80.354$ GeV is the SM prediction
 for the SM  reference point, $(\mt, \mh,
\Delta \alpha_{\rm had}^{(5)}(\mzsq))
=(172 {\rm GeV}, 120 {\rm GeV}, 0.0277)$.
 The dependences of the SM prediction on $m_t$ and
 $\Delta \alpha_{\rm had}^{(5)}(\mzsq)$ are shown at the
 bottom.  Also shown as the red/solid line and the yellow/shaded band
 are the experimental
 mean and the 1-$\sigma$ uncertainty, respectively.}
\label{fig:mw:susy}
\end{figure}
%%%---------------------------------------------------------------

In our framework, the $W$ boson mass is a quantity which can be
calculated from input parameters.
The predicted SM value of $m_W$ for our SM reference point
$(m_t, \mh, \Delta \alpha_{\rm had}^{(5)}(m_Z^2))
 =(172 {\rm GeV}, 120 {\rm GeV}, 0.0277)$
is given in Fig.~\ref{fig:mw:susy} as
the vertical dashed line at $m_W= 80.354$ GeV. 
We see that it is away from the experimental result, whose mean
is shown as the solid vertical line at $m_W=80.399$ GeV together with
its uncertainty shown as the band, roughly at 2-$\sigma$ level.
At the bottom of the figure we show the dependences of the 
SM prediction for $m_W$ on $m_t$ and
$\Delta \alpha^{(5)}_{\rm had}(\mzsq)$.
When $\mt$ becomes larger the prediction also becomes larger because
$\Delta m_W$ has a rather strong dependence on $\Delta T$, see
Eq.~(\ref{eq:Delta_m_W}), and $\Delta T$ has also positive dependence on
$\mt$ as Eq.~(\ref{eq:DTZ_SM}).  The dependence on
$\Delta \alpha_{\rm had}^{(5)}(\mzsq)$ is not as strong as on $m_t$,
but is not negligible.

In Fig.~\ref{fig:mw:susy}, we also show the individual
contributions to the $W$ boson mass from each sector in the MSSM. 
In the figure we take $\tan\beta=10$, but these SUSY contributions
do not change very much for $10\ltsim \tan\beta \ltsim 50$.
The squark and slepton contributions to $\mw$ are examined for the same
parameter space in Fig.~\ref{fig:oblique:tanb10}. They make the fit to
the $\mw$ data better than the SM.  
The sfermion contributions shift
$\mw$ into the 1-$\sigma$ favored range,
when $m_{\wt{Q}} \simlt 500 {\rm GeV}$ or when
$m_{\wt{L}} \simlt 140 {\rm GeV}$.  These improvements mainly come
from the $\Delta T$ and the $\Delta S$ terms in Eq.~(\ref{eq:Delta_m_W})
for the squark and the slepton, respectively.
As for the dependence on the $A_{\rm eff}$ term, 
in the case of squarks, a larger left-right mixing
makes the correction to $\mw$ smaller. 
This can also be explained by the dependence of $\Delta T$
on $A_{\rm eff}$, as already seen in Fig.~\ref{fig:oblique:tanb10}.
Similarly, also for the sleptons, when the left-right 
mixing is not extremely large, the larger $A_{\rm eff}$
predicts smaller $\Delta m_W$.  However, when the left-right 
mixing is extremely large, the contribution to $\Delta T$ becomes
large, as already discussed, which also makes $\Delta m_W$ large,
as seen for $(m_L,m_{\tilde{\tau}_1})=(300,\sim100)$ GeV
in Fig.~\ref{fig:mw:susy}.

The ino contributions to $\mw$ are examined for 
$110 {\rm GeV} \le m_{\wt{\chi}^-_1} \le 140 {\rm GeV}$
and for $M_2/\mu=0.01, 1$
and 100.  They are relatively small compared to the
squarks and sleptons.  Among the three cases,
only the mixed case $(M_2/\mu=1)$ gives a sizable
correction to $\mw$.  This can also be understood from the 
discussion on $\Delta S$ and $\Delta T$ in Section 4.1.

In Fig.~\ref{fig:mw:susy} we also show the predictions from 
the sample SUSY parameter sets. We see that for all the SUSY sample
points the predicted values for $\mw$ are improved compared to the 
SM reference point.  Among them, when there is a light slepton,
like at MSSM2, MSSM4 and MSSM6, the improvement is large
since the slepton contributions are larger than inos for 
similar masses.
In particular, at MSSM2,
where both the sleptons are light, the improvement is most effective.

Also for the predictions from the selected SUSY breaking
scenarios, SG1,\ldots,MM3, the points with light sleptons
make large contributions to $m_W$, like SG1, GM2, MM1--3.
At SG2, the left-right mixing of the stau makes the contribution
large.  At GM1, since there are no light sleptons, the 
contribution is small.

In Fig.~\ref{fig:mw:susy} we do not show the dependence
on the Higgs sector since it is known to be small~\cite{CH2000ew}.
For our selected scenarios SG1,\ldots,MM3, the contributions
from the Higgs sector is $-3{\rm MeV} \lsim \Delta m_W<0$,
which is negligible compared to the experimental uncertainty.

%---------------------------------------------------
\subsection{$\Gamma_W$ in the MSSM}
%---------------------------------------------------

The SUSY corrections to the $W$ boson decay width, $\Gamma_W$, can also
be calculated once the SUSY parameters are fixed. 
The SM prediction 
for $(\mh, m_t)=(120,172)$ GeV is $\Gamma_W =2.090$ GeV,
which is consistent with the experimental value,
$\Gamma_W =2.085 \pm 0.042$ GeV~\cite{PDG10}.
Compared to the experimental uncertainty, the SUSY corrections to 
$\Gamma_W$ are very small ($\sim 0.001$ GeV) for our sample SUSY 
parameters, and we find that $\Gamma_W$ is not as useful as
other EW precision parameters to constrain SUSY contributions.

\subsection{Summary of Electroweak Observables}

\begin{table}
  \rotatebox{90}{ \begin{minipage}{\textheight}
 \begin{center}
\begin{tabular}{|r|c|c|c|c|c|c|c|c|}
\hline
       & data   & SM     
 & MSSM1 & MSSM2 & MSSM3 & MSSM4 & MSSM5 & MSSM6 \\
\hline
$\Delta S$ &&&
  $\hph0.033$ & $-0.025$ & $-0.001$ & $-0.010$ & $\hph$0.009 & $\hph$0.029\\
$\Delta T$ &&&
  $\hph0.040$&$\hph0.048$&$\hph0.016$&$\hph0.023$ &$\hph$0.005& $\hph$0.038 \\
$\Delta R_Z$ &&&
  $-0.032$ & $-0.017$     & $-0.023$ & $-0.007$ &    $-0.020$ & $-0.002$\\
$\Delta m_W$ (GeV) &&&
$\hph$0.017 & $\hph$0.032 & $\hph$0.009 & $\hph$0.013 &  
               $\hph$0.002 & $\hph$0.009 \\ \hline
\hline
$\Gamma_Z$ (GeV)& $\gammazdata$ &   2.4948 &
  2.4954 & 2.4960 & 2.4945 & 2.4952 &               2.4943 & 2.4950\\
$\sigma^0_h$ (nb)& $\sigmahdata$ & 41.481 & 
  41.494 & 41.477 & 41.488 & 41.481 &                41.486 & 41.482\\
$R_l$ & $\rldata$      & 20.737  & 
  20.734 & 20.745 & 20.731 & 20.739 &                  20.733 & 20.735\\
$A^{0,l}_{\rm FB}$ & $\afbldata$ & 0.01613 &
 0.01626 & 0.01651 & 0.01622 & 0.01628 &              0.01614 & 0.01611\\
\hline
$R_b$ & $\rbdata$   & 0.21585 & 
  0.21587& 0.21586& 0.21586 & 0.21585 &                0.21578 & 0.21578\\
$R_c$ & $\rcdata$   & 0.1722$\hpz$  & 
  0.1722 & 0.1722 & 0.1722 & 0.1722 &              0.1722 & 0.1722\\
$\ast$ \hspace{0.73cm} $A^{0,b}_{\rm FB}$ & $\afbbdata$ &  0.1028$\hpz$ & 
  0.1032 & 0.104 & 0.1031 & 0.1033 &                 0.1029 & 0.1028\\
$\ast$ \hspace{0.73cm} $A^{0,c}_{\rm FB}$ & $\afbcdata$ &  0.0735$\hpz$ &
  0.0738 & 0.0744 & 0.0736 & 0.0738 &                  0.0734 & 0.0734\\
$\ast$ \hspace{0.0cm}  
$\sin^2 \theta_{\rm eff}^{\rm lept}$ & $\jetdata$  & 0.2316$\hpz$  &
  0.2315 & 0.2314 & 0.2315 & 0.2315 &                 0.2316 & 0.2316\\
$\ast$ \hspace{1.01cm}
$A_b$ & $\abdata$ & 0.935$\hpzz$  &
  0.9347 & 0.9348 & 0.9346 & 0.9347 &                 0.9354 & 0.9354\\
$\ast$ \hspace{1.01cm}
$A_c$ & $\acdata$ & 0.668$\hpzz$  &
   0.668 & 0.669 & 0.668 & 0.668 &                    0.668 & 0.668 \\
\hline
$A_\tau(P_\tau)$ & $\altaupoldata$  & 0.1467 &
  0.1473 & 0.1483 & 0.1471 & 0.1473 &                   0.1467 & 0.1466\\
$A_{\rm LR}^0(A_e)$ & $\alSLDdata$  & 0.1467 &
  0.1473 & 0.1483 & 0.1471 & 0.1473 &                   0.1467 & 0.1466\\
\hline
$m_W$ (GeV)& $\mwdata$          & 80.354 & 
  80.370 & 80.386 & 80.363 & 80.367 &                    80.356 & 80.362\\
$\Gamma_W$ (GeV)& $\hpz\gammawdata$  & $\hpz$2.090 & 
 2.092 & 2.093 & 2.091 & 2.091 &                  2.091 & 2.091 \\
\hline \hline
$\chi^2_{\rm EW}$ (all) &                       & 20.61 &
 18.16 & 18.37    & 20.07 & 19.21 &                  21.74 & 20.39 \\
$\chi^2_{\rm EW}$ (excl. $A^{0,b}_{\rm FB}$) &  & 16.32 &
 12.76 & 10.59    & 15.08 & 13.71 &                  17.26 & 16.11 \\
$\chi^2_{\rm EW}$ (excl. *) &                   & 14.83 &
 11.02 & \hpz8.30 & 13.43 & 11.94 &                  15.72 & 14.61 \\
\hline
\end{tabular} \end{center}
 \caption{{\small
 The breakdown of the radiative corrections at our SUSY sample points
 MSSM1 -- MSSM 6.
 In the column labeled as ``SM'', the values for the SM reference
 point 
$(m_t, \mh, \Delta \alpha_{\rm had}^{(5)}(m_Z^2))
 =(172 {\rm GeV}, 120 {\rm GeV}, 0.0277)$ are given.
 The values for the SUSY corrections $\Delta S$, $\Delta T$,
 $\Delta R_Z$ and $\Delta m_W$
 are the deviations from this SM reference point.}}
 \label{tab:SUSYbreakdown}
\end{minipage} }
\end{table}

\begin{table}[phtbc]
  \rotatebox{90}{ \begin{minipage}{\textheight}
 \begin{center}
  \begin{tabular}{|r|c|c|c|c|c|c|c|}
   \hline
       & SG1 & SG2 & GM1 &GM2 &MM1  & MM2 & MM3 \\
   \hline
   $\Delta S^{}$ 
                   & $\hph$0.008 & $\hph$0.012 & $\hph$0.009
                   & $\hph$0.006 & $\hph$0.003 & $\hph$0.005 & $\hph$0.003 \\
   $\Delta T^{}$ 
                   & $\hph$0.044 & $\hph$0.054 & $\hph$0.011
                   & $\hph$0.028 & $\hph$0.034 & $\hph$0.019 & $\hph$0.027 \\
   $\Delta R_Z^{}$
                   & $-0.019$    & $-0.002$    & $-0.005$
                   & $-0.009$    & $-0.011$    & $-0.011$ & $-0.019$ \\ 
   $\Delta m_W$ (GeV)
                   & $\hph$0.019 & $\hph$0.020 & $\hph$0.003
                   & $\hph$0.013 & $\hph$0.015 & $\hph$0.008 & $\hph$0.013 \\
   \hline \hline
   $\Gamma_Z^{}$ (GeV) &
       2.4958 & 2.4957  & 2.4948  & 2.4955  & 2.4956 & 2.4952 & 2.4952 \\ 
   $\sigma^0_h$ (nb)   &
      41.486  & 41.484  & 41.485  & 41.483  & 41.483 & 41.485 & 41.486 \\ 
   $R_l$        &
      20.738  & 20.734  & 20.734  & 20.739  & 20.739 & 20.736 & 20.736\\ 
   $A^{0,l}_{\rm FB}$  &
      0.01631 & 0.01630 & 0.01615 & 0.01626 & 0.01629 & 0.01620 & 0.01627\\
     \hline
   $R_b$ & 
     0.21589 & 0.21569 & 0.21575 & 0.21588 & 0.21588 & 0.21587 & 0.21588\\
   $R_c$ & 
     0.1722$\hpz$  & 0.1723$\hpz$  & 0.1722$\hpz$  &
     0.1722$\hpz$  & 0.1722$\hpz$  & 0.1722$\hpz$  & 0.1722$\hpz$ \\
$\ast$ \hspace{0.73cm}   $A^{0,b}_{\rm FB}$  &
     0.1034$\hpz$  & 0.1035$\hpz$  & 0.1030$\hpz$  & 
     0.1032$\hpz$  & 0.1033$\hpz$  & 0.1030$\hpz$  & 0.1032$\hpz$ \\
$\ast$ \hspace{0.73cm}   $A^{0,c}_{\rm FB}$  &
     0.0739$\hpz$  & 0.0739$\hpz$  & 0.0735$\hpz$  &
     0.0738$\hpz$  & 0.0738$\hpz$  & 0.0736$\hpz$  & 0.0738$\hpz$\\
$\ast$ \hspace{0.0cm}  $\sin^2 \theta^{\rm lept}_{\rm eff}$      &
     0.2315$\hpz$  & 0.2315$\hpz$  & 0.2316$\hpz$  &
     0.2315$\hpz$  & 0.2315$\hpz$  & 0.2315$\hpz$  & 0.2315$\hpz$ \\ 
$\ast$ \hspace{1.01cm} $A_{b}$ & 
     0.935$\hpzz$  & 0.936$\hpzz$  & 0.936$\hpzz$   
   & 0.935$\hpzz$  & 0.935$\hpzz$  & 0.935$\hpzz$  & 0.935$\hpz$ \\
$\ast$ \hspace{1.01cm} $A_{c}$ & 
     0.668$\hpzz$  & 0.668$\hpzz$  & 0.668$\hpzz$   
   & 0.668$\hpzz$  & 0.668$\hpzz$  & 0.668$\hpzz$  & 0.668$\hpz$ \\
\hline
%---------
$A_\tau(P_{\tau})$  &
          0.1475   & 0.1474  & 0.1468 & 0.1472  & 0.1474 & 0.1470 & 0.1473\\
$A^0_{LR} (A_e)$    &
          0.1475   & 0.1474  & 0.1468 & 0.1472  & 0.1474 & 0.1470 & 0.1473\\
\hline
%---------
$m^{}_W$ (GeV)      &
     80.373  & 80.373  & 80.357  & 80.366 & 80.369  & 80.362 & 80.366 \\
%---------
$\Gamma^{}_W$ (GeV) &
    $\hpz$2.092 & $\hpz$2.091 & $\hpz$2.090 &
    $\hpz$2.091 & $\hpz$2.091 & $\hpz$2.091 & $\hpz$2.092 \\
\hline \hline
%---------
$\chi^2_{\rm EW}$ (all)            &
     18.08 & 19.38 & 21.44 & 18.97 & 18.65 & 19.70 & 19.05\\
$\chi^2_{\rm EW}$ (excl. $A^{0,b}_{\rm FB}$) &    
     12.19 & 13.23 & 16.69 & 13.53 & 13.01 & 14.85 & 13.61\\
$\chi^2_{\rm EW}$ (excl. $\ast$)  &
     10.34 & 11.30 & 15.08 & 11.78 & 11.22 & 13.24 & 11.86\\
\hline
  \end{tabular} 
 \caption{{\small
 The breakdown of the radiative corrections at our sample points
in the selected SUSY models.}}
 \label{tab:SUSYbreakdown2}
\end{center}
  \end{minipage} }
\end{table}

In Tables~\ref{tab:SUSYbreakdown} and
\ref{tab:SUSYbreakdown2} we summarize the values of the
SUSY contribution to the oblique parameters and the electroweak
observables for our sample parameters.  From the tables, we see
that for our sample parameters the SUSY corrections are small in 
general since for those points the SUSY particles are at
the range of a few hundred GeV or heavier. 

Next, if we look into the observables in 
Tables~\ref{tab:SUSYbreakdown} and \ref{tab:SUSYbreakdown2},
we see that the observables like
$A_c$ and $\sin^2\theta_{\rm eff}^{\rm lept}$  
do not depend on SUSY parameters very much compared
to the experimental accuracy.
We also see that some jet asymmetry observables like
$A_{\rm FB}^{0,b}$, $A_{\rm FB}^{0,c}$ and $A_b$ do not agree between
experiment and SM, which SUSY contributions cannot improve very much,
as is well known. 

We also give $\chi^2_{\rm min}$ for the cases where
(i) all the data are used, (ii) only $A_{\rm FB}^{0,b}$ 
is excluded, (iii) the jet asymmetry data (data with
$\ast$ in the tables) are excluded.  We see that 
the values of $\chi^2_{\rm min}$ show sizable changes
between the cases (i) and (ii), but not very much between
(ii) and (iii).  
This may suggest that 
$A_{\rm FB}^{0,b}$ is the main source of the
deviation of the SM from the data.

In Tables~\ref{tab:SUSYbreakdown} and \ref{tab:SUSYbreakdown2} 
we also give the SUSY contribution to
the shift $\Delta m_W$. Since the shift can be written in terms of the
oblique parameters by Eq.~(\ref{eq:Delta_m_W}), we can calculate the
shift also from the values of $\Delta S$, $\Delta T$, and so on in
Tables~\ref{tab:SUSYbreakdown} and
\ref{tab:SUSYbreakdown2}.  Since for our sample points $\Delta T$
is larger than $\Delta S$, $\Delta U$ and 
$-\Delta \bar{\delta}_G/\alpha$, $\Delta T$ gives the main contribution
to $\Delta \mw$. In those sample points with larger $\Delta T$, such as
SG1 and SG2, the predicted shift $\Delta \mw$ is
larger, which makes the fit of $\mw$ better. Similarly, for those point
with smaller $\Delta T$, such as GM1 and MM2, the shift $\Delta \mw$ is
small, which makes the total $\chi^2$ worse.

The SUSY contribution to $\Gamma_W$ is small
compared to the experimental accuracy, 
as seen from Tables~\ref{tab:SUSYbreakdown} and \ref{tab:SUSYbreakdown2}.
We conclude that it is not very useful to constrain SUSY contributions.

In this paper, we do not consider the SUSY non-oblique corrections
other than $\Delta \overline{\delta}_G$.
In the MSSM we expect that the corrections to 
the $Z$-$b$-$\bar{b}$ vertex is the largest among vertex corrections.
We find that the SUSY contributions to
$\Delta g_{L/R}^b$ are at most of the order of $10^{-4}$,
which is far smaller compared to the oblique corrections.

\newpage

\section{Preferred parameters in a few SUSY breaking models}

In the previous section, we have studied the 
constraints from the EW precision data
on $\Delta S_Z$, $\Delta T_Z$, and $m_W$.
In Figs.~\ref{fig:oblique:tanb10} and \ref{fig:obl_nobjet}, 
the constraints on $\Delta S_Z$ and $\Delta T_Z$ are shown
by the favored region of an elliptic shape with a strong positive
correlation.  This means that the linear combination of $\Delta S_Z$
and $\Delta T_Z$ along the minor axis of the ellipse is constrained
much stronger than the orthogonal combination along the major
axis.  We also note that this combination along the minor axis
direction is strongly affected by the removal of the jet
asymmetry constraints.  In this section, we show the constraints
on our $g-2$ favored SUSY models from the electroweak precision
measurements in the plane of this strongly constrained combination
and $m_W$, as two-dimensional `summary plots'.  For those models
of SUSY breaking where the squark masses are related to the
slepton and ino particle masses, we also examine the constraints
in the plane of muon $g-2$ and ${\rm Br}(b\to s \gamma)$.  
The MSSM model points do not appear in those plots since squark
masses can be set large to make them consistent with the 
${\rm Br}(b\to s \gamma)$ constraint.

\begin{figure}
\begin{center}
\includegraphics[angle=270,scale=0.5]{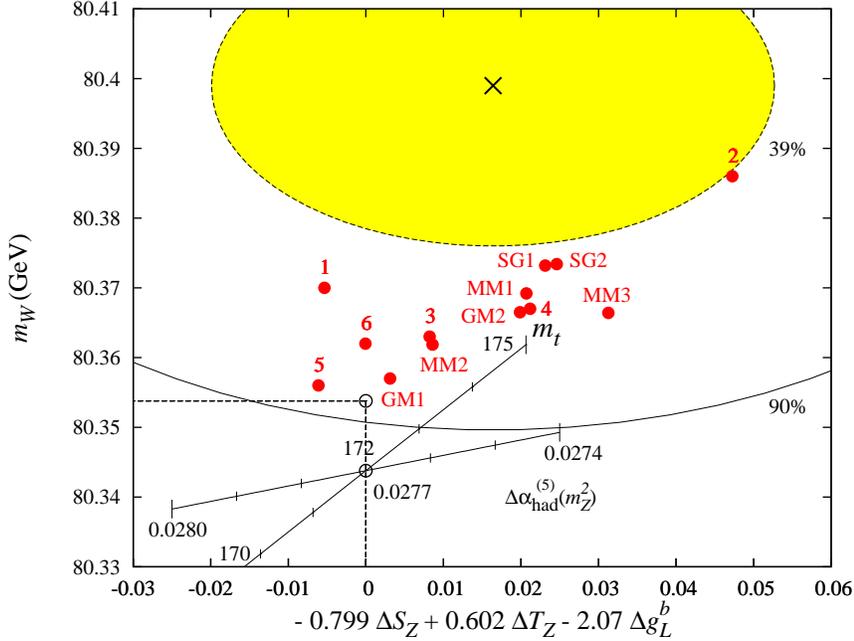}
\end{center}
\vspace*{-0.5cm}
\caption{\label{fig:STmW}
The favored region in the plane of the most experimentally
 constrained linear combination of $\Delta S_Z$ and $\Delta T_Z$ versus
 $\mw$. The inner/outer ellipses stand for the 39\%/90\% CL
 favored regions.  The upper open circle is the 
 SM prediction for $(m_t, \mh, \Delta \alpha_{\rm had}^{(5)}(m_Z^2))
 =(172 {\rm GeV}, 120 {\rm GeV}, 0.0277)$.
 Also shown as the red/filled blobs are the
 predictions of our sample SUSY models.
 The lines with ticks around the lower open circle are
 the ``rulers'' to show how the SM prediction, and hence
 all the SUSY model prediction points as well, shifts when
 more accurate data on $m_t$ and $\Delta \alpha_{\rm had}^{(5)}(m_Z^2)$
 are available.}
\end{figure}

In the analysis we have performed in Fig.~\ref{fig:oblique:tanb10},
the direction of the minor axis of the ellipse, which corresponds to 
the most tightly constrained direction, is
$-0.799 \Delta S_Z +0.602 \Delta T_Z$, along which 
\begin{align}
 - 0.799 \Delta S_Z +0.602 \Delta T_Z
 = 0.016 + 2.07 \Delta g_L^b + 0.025 \xa \pm 0.036.
\end{align}
We can conveniently combine this result with the constraint from $m_W$
in a single figure (Fig.~\ref{fig:STmW}).   In the figure, 
the 39\% and the 90\% CL favored regions are shown as the ellipses.
Also shown as the upper open circle is the SM prediction for
our reference point,  $(m_t, \mh, \Delta\alpha_{\rm had}^{(5)}(m_Z^2))
=(172 {\rm GeV}, 120 {\rm GeV}, 0.0277)$.
The SM predictions for different $\mt$ within
$169{\rm GeV} \le m_t \le 175{\rm GeV}$ can be read off using the
``ruler'' around the lower open circle.  For example, if we
take $\mt=175 {\rm GeV}$ instead of $\mt=172 {\rm GeV}$, the SM prediction
moves from the upper open circle toward the upper-right,
in the direction of the vector whose initial point is the lower open circle
with the terminal point being the point shown as ``175''
and also by the length of the same vector.
We see that, within the range of the top quark mass shown,
a larger top quark mass is favored from the data.  This preference
for a larger top quark mass is clearer in $m_W$ than in 
the most constrained combination of  $\Delta S_Z$-$\Delta T_Z$.

In Fig.~\ref{fig:STmW}, we also plot the sample SUSY points.
In the direction of the most
constrained $\Delta S_Z$-$\Delta T_Z$ combination the fit does not
improve very much by introducing SUSY particles since the SM already
gives a good description.  It is also seen that all our SUSY sample
points lie within the 1-$\sigma$ favored range of this most constrained
direction.  These tendencies could already be read off in
Fig.~\ref{fig:oblique:tanb10}.  In the $m_W$ direction, we have
improvements in general as discussed in Section 4.3.

\begin{figure}
\begin{center}
\includegraphics[angle=270,scale=0.5]{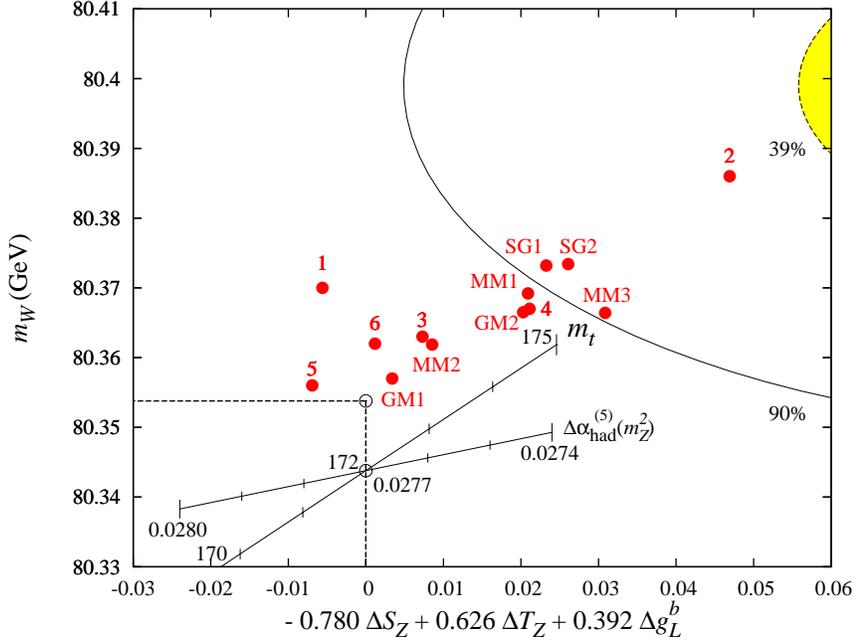}
\end{center}
\vspace*{-0.5cm}
\caption{The same as Fig.~\ref{fig:STmW} except that
the jet asymmetry data are not included in the analysis.}
\label{fig:STmW_woja}
\end{figure}

We can repeat the same analysis as Fig.~\ref{fig:STmW} also in the
case where the jet asymmetry data are not included in the input data. 
The most constrained direction in the 
$\Delta S_Z$-$\Delta T_Z$ plane in this case is
\begin{align}
 -0.780 \Delta S_Z + 0.626 \Delta T_Z = 0.100
 - 0.39 \Delta g_L^b + 0.024 \xa \pm 0.044.
\end{align}
Combining this with the constraint from $m_W$, 
we show the favored region in Fig.~\ref{fig:STmW_woja}. 
Compared to Fig.~\ref{fig:STmW}, we see that 
the ellipses move toward the right, 
which is the negative $\Delta S_Z$ direction.
We also show the SM prediction for the reference point as the
upper open circle, which will move according to the ``ruler'' 
around the lower open circle for a different $m_t$.
In this case, there is a clearer tendency that the large
top quark mass is favored within the range of $m_t$ shown.
We also show the predictions from our sample SUSY points.
As a general
tendency, our SUSY scenarios slightly improve the fit over the SM
reference point. In this case, the light slepton scenario,
MSSM2, can improve the fit most efficiently among our sample
SUSY points, since we have chosen the slepton mass small in such a way
that it can better explain the negative $\Delta S_Z$. The degree of the
improvement can also be seen in the $\chi^2$ without jet asymmetry data
of Tables~\ref{tab:SUSYbreakdown} and
\ref{tab:SUSYbreakdown2}.  At MSSM2,
$\chi^2_{\rm min}=8.30$, which is much better than
that of the SM reference point,
$\chi^2_{\rm min}=14.83$, and also than other sample SUSY points.

Having discussed the EW constraints,
we now take the constraint from $b\to s \gamma$ into account.
The experimental value quoted in RPP 2010~\cite{PDG10} is
${\rm Br}(B \to X_s \gamma) = (3.55 \pm 0.26) \times 10^{-4}$,
while the SM prediction at NNLO is 
${\rm Br}(b \to s \gamma) = (3.15 \pm 0.23) \times 10^{-4}$
in Ref.~\cite{Misiak:2006zs},
and ${\rm Br}(b \to s \gamma) = (2.98 \pm 0.26) \times 10^{-4}$
in Ref.~\cite{Becher:2006pu}.
Since the experimental and the theoretical values
agree within 2-$\sigma$ level,
it is preferred that the SUSY contribution is not very 
large so that it does not spoil the rough agreement.

To suppress the SUSY contributions, we can think of two possibilities:
either the relevant SUSY particles are heavy enough, or
a cancellation among the relevant diagrams happens. 
To see how this can be realized, let us look into the structure
of the SUSY contributions.
At one-loop, the SUSY contribution mainly comes from the 
chargino--stop loop and the charged-Higgs--top loop diagrams.
We neglect the possible contributions from the 
gluino--sbottom loop diagram, assuming
the minimal flavor violation~\cite{Gabrielli:1994ff}.
Under this assumption, 
we are only interested in those parameter sets in which the
chargino--stop and the charged-Higgs--top contributions
cancel with each other to the extent that the experimental
constraint is satisfied, or those parameter sets in which 
the relevant SUSY particles are heavy enough.

\begin{figure}
\begin{center}
\includegraphics[height=8cm]{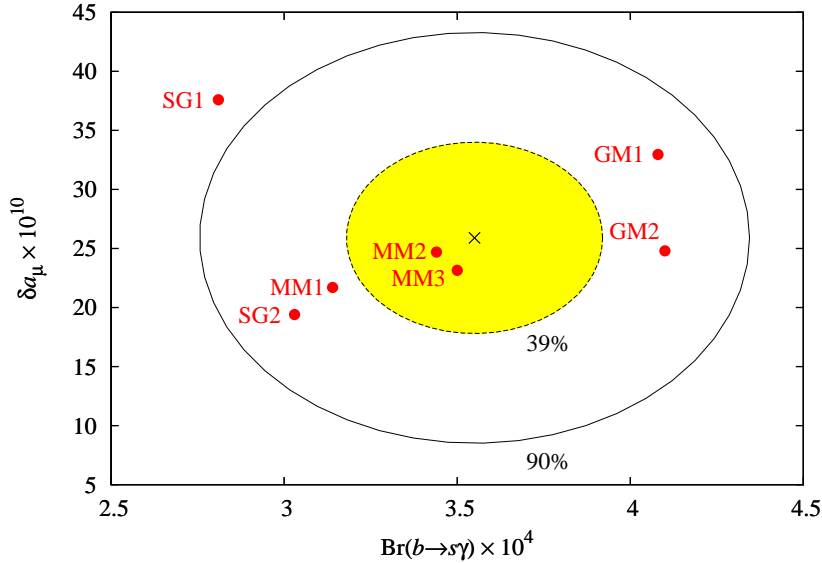}
\end{center}
\vspace*{-0.7cm}
\caption{The branching ratio of $b\to s\gamma$ and 
the SUSY contribution to the muon $g-2$ at our sample points.
The inner and the outer ellipses stand for the 39\% and the
90\% CL favored regions, respectively.}
\label{fig:ellipse}
\end{figure}

In Fig.~\ref{fig:ellipse} we show the predictions for
the muon $g-2$ and Br$(b \to s \gamma)$ for these sample points.
Also shown as the inner and the outer ellipses are 
the 39\% and the 90\% CL contours, respectively.
In the figure, to specify the favored region of ${\rm Br}(b \to s \gamma)$, 
we use the experimental result
${\rm Br}(B \to X_s \gamma) = (3.55 \pm 0.26) \times 10^{-4}$.
As for the uncertainty in the Standard Model prediction, we
assign $0.26\times 10^{-4}$, which is the larger of the 
uncertainties in the two SM predictions mentions above.
We add the uncertainties in the experimental
results and the SM prediction in quadrature.
Concerning the MSSM prediction for ${\rm Br}(b\to s\gamma)$,
we use {\tt micromegas} version 2.0.7~\cite{micromegas}, 
while the SUSY contribution to the muon $g-2$ is calculated
by using our own code.

From the figure, concerning ${\rm Br}(b\to s\gamma)$,
we see that all the points are within
the 90\% CL favored region.  We also see that 
${\rm Br}(b\to s\gamma)$ at SG1 is a little bit farther 
from the central point than the other points.
This happens since, at SG1, the third generation squarks and
the lighter charginos are slightly lighter than those of the other
points, and since the cancellation among the SUSY diagrams 
are milder.

Concerning the muon $g-2$ for those sample points, since we
have already discussed in Section 2, we do not repeat it here.

As an additional constraint on these SUSY sample points,
we now comment on the dark matter relic density predicted from
these models.  At all our SUSY sample parameters, 
the lightest SUSY particle (LSP) is stable
because of the $R$-parity conservation, and hence 
the LSP is a potential candidate for dark matter.
The LSP is the lightest neutralino
in our sample points based on mSUGRA or the mirage 
mediation (MM), while those based on 
the gauge-mediated models the LSP is gravitino.
In either cases, the relic density of the LSP is calculable.
For the mSUGRA and MM based points, 
we have calculated the LSP relic density
using {\tt micromegas}.   
The results are
$\Omega h^2$ = 0.08, 0.01, 0.11, 0.08 and 0.001 for SG1,
SG2, MM1, MM2 and MM3, respectively.  For all these
cases except MM3, the LSP is an almost pure bino, with a very small
mixture from Higgsinos and wino.  For MM3, since the LSP
is wino, the relic density is smaller because of the 
larger annihilation cross section of the wino LSP. 
As for the gauge
mediated model sample points, GM1 and GM2, the 
LSP mass, namely the gravitino mass, is in the eV range,
in which case the LSP relic density is negligible.
These relic densities should be compared to the results
of a global fit of the cosmological parameters on  
the non-baryonic matter density $\Omega_{\rm DM} h^2$~\cite{PDG10},
\begin{eqnarray}
\Omega_{\rm DM} h^2 = 0.110\pm 0.006.
\label{eq:omega_DM}
\end{eqnarray}
We see that for all our sample points, the relic density
of the LSP is nearly equal to or less than the observed density
of dark matter.  If the relic density of the LSP were 
significantly larger than the value in Eq.~(\ref{eq:omega_DM}),
those models would be excluded.  On the contrary, if the relic
density of the LSP is less than the value in Eq.~(\ref{eq:omega_DM}),
such a model can still be phenomenologically 
viable since there is still a possibility that an unknown
particle like an axion can also contribute to the dark matter
density.  Hence we conclude that our sample parameters 
are not excluded from the dark matter density calculations.

We do not include constraints from the
low-$Q^2$ precision measurements such as atomic parity
violation and neutrino-nucleon scatterings at low energies
since constraints from these measurements are known to be 
much less stringent than those from the $Z$-pole experiments.
Another class of observables we do not include is those from 
$B$-physics, namely ${\rm Br}(B^0_s \to \mu^+\mu^-)$,
${\rm Br}(B^+ \to \tau^+ \nu_\tau)$ and $\Delta m_s$.
Even though these observables potentially give 
non-trivial constraints on large $\tan\beta$
models~\cite{Bona:2009cj}, we do not include them
since our main interest in the present paper is in 
the signal from the slepton and the ino sectors rather than 
the squark and the Higgs sectors.

\section{Summary}

We have studied impacts of recent muon $g-2$ measurements and
the LEP final electroweak data on the MSSM.
We identify several regions of the MSSM parameter space which
fill the gap between the SM prediction and the observed value
of the muon $g-2$, and at the same time have observable effects
for the electroweak precision measurements.  In all the
selected regions of the MSSM parameter space, the MSSM
predictions are consistent with the LEP/SLC $Z$ boson observables,
while improve the SM fit to the $W$ boson mass slightly.
When we remove the constraints from the jet asymmetry measurements
at LEP/SLC, we find that MSSM models with very light sleptons
($\ltsim$ 200 GeV) and
moderately heavy ino particles ($\sim$ several 100 GeV) are
favored over models with a very light chargino ($\sim$ 100 GeV)
and moderately light sleptons ($\sim$ a few 100 GeV).

We also examined a few models of SUSY breaking scenarios,
including minimal SUGRA models, gauge mediation models,
and the mixed moduli and anomaly mediation models.
All of them have parameter region with relatively light
slepton and ino particles which contribute to the muon $g-2$.
Those models with moderately heavy smuons and ino
particles can still contribute to the muon $g-2$ with
large $\tan\beta$ ($\gtsim$ 40), and can at the same time improve
the fit to $m_W$ and the $Z$ boson parameters if there is a
significant mixing in the stau sector.  Sample
scenarios in each SUSY breaking models are found which
improves the SM fit to the muon $g-2$, $m_W$, the $Z$
parameters in the absence of jet asymmetry data, and are still
compatible with ${\rm Br}(b\to s \gamma)$.
We believe that our investigations will help us identifying
the supersymmetry breaking scenario once signatures of
SUSY particle productions are discovered at the LHC.

\section*{Acknowledgements}

We thank Y.~Shimizu for providing us with the SUSY parameters
for the sample point GM1, and A.~Crivellin, J.~Girrbach
and U.~Nierste for comments on the large $A$-term scenario
and on the effect of $\tan\beta$-enhanced resummation. 
KH wishes to thank Aspen Center
for Physics where stimulating discussions with
the participants of the 2008 summer program took place.
DN would like to thank K.~Okumura for useful discussions
on the mirage mediation models.
This work is supported in part by Grants-in-Aid for Scientific
Research (No.\ 18340060 and 20340064) from the Japan Society for
the Promotion of Science (JSPS).

%%%%%%%%%%%%%%%%%%%%%%%%%%%%%%%%%%%%%%%%%%%%%%
%%%%%%%                                %%%%%%%
%%%%%%%	        Journal code           %%%%%%%
%%%%%%%                                %%%%%%%
%%%%%%%%%%%%%%%%%%%%%%%%%%%%%%%%%%%%%%%%%%%%%%
\def\PRD#1#2#3{Phys. Rev. {\bf D#1}, #3 (#2)}
\def\NATURE#1#2#3{Nature {\bf #1}, #3 (#2)}
\def\NPB#1#2#3{Nucl. Phys. {\bf B#1}, #3 (#2)}
\def\ZPC#1#2#3{Z. Phys. {\bf C#1}, #3 (#2)}
\def\EPJC#1#2#3{Eur. Phys. J. {\bf C#1}, #3 (#2)}
\def\PRL#1#2#3{Phys. Rev. Lett. {\bf #1}, #3 (#2)}
\def\PLB#1#2#3{Phys. Lett. {\bf B#1}, #3 (#2)}
\def\JHEP#1#2#3{JHEP {\bf #1}, #3 (#2)}

\end{document}